\documentstyle[11pt,aaspp4]{article}

\newcommand{\mic}{$\mu$m}

\begin{document}
\title{THE COBE DIFFUSE INFRARED BACKGROUND EXPERIMENT SEARCH FOR THE COSMIC INFRARED BACKGROUND: IV. COSMOLOGICAL
IMPLICATIONS}
\author{E. Dwek\altaffilmark{1}, R. G. Arendt\altaffilmark{2}, M. G. Hauser\altaffilmark{3},
D. Fixsen\altaffilmark{2}, T. Kelsall\altaffilmark{1}, D. Leisawitz\altaffilmark{4}, Y. C. Pei\altaffilmark{3}, E. L.
Wright\altaffilmark{5},  
 J. C. Mather\altaffilmark{1}, S. H. Moseley\altaffilmark{1},
 N. Odegard\altaffilmark{2}, R. Shafer\altaffilmark{1}, R. F. Silverberg\altaffilmark{1}, and J. L.
Weiland\altaffilmark{2}}

\altaffiltext{1}{Laboratory for Astronomy and Solar Physics, Code 685,  NASA/Goddard
Space Flight Center, Greenbelt, MD 20771.\ \ eli.dwek@gsfc.nasa.gov}
\altaffiltext{2}{ Raytheon STX, Code 685, NASA/GSFC, Greenbelt, MD 20771}
\altaffiltext{3}{Space Telescope Science Institute, 3700 San Martin Drive, Baltimore, MD
21218}
\altaffiltext{4}{Laboratory for Astronomy and Solar Physics, Code 681,  NASA/Goddard
Space Flight Center, Greenbelt, MD 20771.}
\altaffiltext{5}{UCLA, Astronomy Department, Los Angeles, CA 90024-1562}

\begin{center}
to be published in The Astrophysical Journal\\
received: February 3, 1998; accepted: June 8, 1998
\end{center}

\begin{abstract} 
A direct measurement of the extragalactic background light (EBL) can provide important
constraints on the integrated cosmological history of star formation, metal and dust production, and
the conversion of starlight into infrared emission by dust. 
In this paper we examine the cosmological implications of the recent detection of the EBL in the 125 to 5000 $\mu$m
wavelength region by the Diffuse Infrared Background Experiment (DIRBE) and Far Infrared Absolute Spectrophotometer (FIRAS) on board
the {\it Cosmic Background Explorer} ({\it COBE}).
We first show that the 140 and 240
$\mu$m isotropic residual emission found in the DIRBE data cannot be produced by foreground emission sources in the solar system or
the Galaxy. The DIRBE 140 and 240 $\mu$m isotropic residuals, and by inference the FIRAS residuals as well, are therefore
extragalactic. Assuming that most
of the 140 and 240
$\mu$m emission is from dust yields a 2$\sigma$ lower limit of
$\nu\ I(\nu) \approx$ 5 nW m$^{-2}$ sr$^{-1}$  for the EBL at 100 $\mu$m.
The integrated EBL detected by the {\it COBE} between 140 and 5000 $\mu$m is $\sim$ 16 nW m$^{-2}$ sr$^{-1}$, roughly 20$-$50\%
 of the integrated EBL
intensity expected from energy release by nucleosynthesis throughout cosmic history. This also implies that at least $\sim$ 5$-$15\% of the
baryonic mass density implied by Big Bang nucleosynthesis has been cycled through stars.
The {\it COBE} observations provide important constraints on the cosmic star formation rate, and we calculate the EBL spectrum 
for various star formation histories. The results show that the UV and optically determined cosmic star formation rates fall short
 in producing the observed 140 to 5000 $\mu$m background. The {\it COBE} observations
 require the star formation rate at redshifts of $z \approx~1.5$ to be larger than that inferred
 from UV$-$optical observations by at least a factor of 2. This excess stellar energy must be mainly generated by massive stars, 
 since it otherwise would result in a local
 luminosity density that is significantly larger than observed. The energy sources could either be yet undetected 
dust$-$enshrouded galaxies, or extremely dusty star$-$forming regions in observed galaxies, and they may be responsible for the 
observed iron enrichment in the intra$-$cluster medium. The exact star formation history 
or scenarios required to produce the 
 EBL at far$-$IR wavelengths cannot be unambiguously resolved by the {\it COBE} observations, and must await future observations.
\end{abstract}

\keywords{cosmology: observations, diffuse radiation - infrared: general, galaxies - galaxies: evolution}

\section{INTRODUCTION}

The extragalactic background light (EBL) consists of the cumulative emission from various
pregalactic objects, protogalaxies and galaxies throughout the evolution of the universe. Its detection is a subject of
great scientific interest and the main purpose of the Diffuse Infrared Background Experiment
(DIRBE) on the {\it Cosmic Background Explorer} ({\it COBE}) spacecraft (Boggess et al. 1992). The DIRBE was designed to
search for a cosmic infrared (IR) background, which is presumed to be spatially isotropic. The detection of the EBL at IR
wavelengths is greatly hampered by the presence of strong foreground emission. The removal of these various layers of
emission without affecting any potential extragalactic component is a formidable task which has been described by the DIRBE
team in three previous papers in this series:  Hauser et al.
(1998; Paper I) summarized limits and detections of the EBL; Kelsall et
al. (1998; Paper II) described the subtraction of the interplanetary dust scattering and emission component; and Arendt et al.
(1998; Paper III) described the subtraction of the Galactic stellar and diffuse interstellar emission components.
A residual may be considered as extragalactic in origin if its signal is positive and isotropic. Hauser et
al. (1998) described in detail the tests conducted to examine whether significant residual emission exists at each of the
DIRBE wavelengths and, if so, whether that emission is isotropic. These studies resulted in upper limits on the EBL in the
1.25 - 100
$\mu$m region, and in the detection of a positive isotropic signal at 140 and 240 $\mu$m.
Furthermore, a
detailed analysis of the data obtained with the Far Infrared Absolute Spectrophotometer (FIRAS) instrument on the {\it COBE}
resulted in a detection of the EBL in the 125 $-$ 5000 $\mu$m region (Fixsen et al. 1998; Paper V). The DIRBE and FIRAS isotropic
residuals can be summarized as:

\begin{eqnarray}
\nu I(\nu) & = & 25.0\ \pm6.9 \ \ \hbox {nW m}^{-2} \hbox {sr}^{-1} \quad \hbox {at 140}\ \mu \hbox{m \quad  (Paper I)}
\nonumber \nl
           & = & 13.6\pm2.5 \ \ \hbox {nW m}^{-2} \hbox {sr}^{-1}\quad \hbox {at 240}\ \mu \hbox{m \quad  (Paper I)} \nl
 & = & (1.3\pm0.4)\times10^{-5}\left({\lambda (\mu m)\over 100}\right)^{-0.64\pm0.12}\ \nu B_{\nu}(18.5\pm1.2\ K) \nonumber
\nl
 &  &   at \quad \lambda = 125-5000\ \mu \hbox{m} \quad \quad \quad \hbox{(Paper V)}, \nonumber
\end{eqnarray}

\noindent
where $B_{\nu}(T)$ is the Planck function at frequency $\nu$ and temperature $T$.
The DIRBE detections are smaller than, but consistent with, the detections reported
by Schlegel, Finkbeiner, \& Davis (1998). The FIRAS detections of Fixsen et al. (1998) are consistent with those of the DIRBE, and, 
at wavelengths longer than 300 $\mu$m, generally similar to
the tentative detections reported by Puget et al. (1996). A more detailed discussion of the comparison 
is given by Hauser et al. (1998) and Fixsen et al. (1998). The integrated intensity of the background detected in the DIRBE
140 and 240 $\mu$m bands is 
10.3 nW m$^{-2}$ sr$^{-1}$, and that detected by the FIRAS in the 240 to 5000 $\mu$m region is 
5.3 nW m$^{-2}$ sr$^{-1}$, giving a total EBL intensity of $\sim$ 16 nW m$^{-2}$ sr$^{-1}$ in the $\sim$ 125 to 5000 $\mu$m 
wavelength interval.

The purpose of this paper is to discuss some of the implications of these {\it COBE} measurements. We show in \S 2 and Appendix $A$  that local
sources in the solar system or the Galaxy cannot provide significant contributions to the residual isotropic emission at 140
and 240
$\mu$m. No known component of the solar system can produce an isotropic emission component at the observed intensity in these
two DIRBE bands. Furthermore, a hypothetical solar system component consistent with the various constraints
provided by the DIRBE observations is unlikely to survive or be maintained over the lifetime of the solar system.
Likewise, we argue that no Galactic dust component can produce the observed far$-$infrared isotropic emission component. The
residual emission detected by the DIRBE and FIRAS instruments is therefore of extragalactic origin. 
Adopting the extragalactic nature of the 140 and 240
$\mu$m residuals, we show that the detections in these bands imply a lower limit on the extragalactic contribution at
100 $\mu$m as well (\S 3).

Various cosmological implications of the {\it COBE} detections are addressed in this paper.

(1) {\it The origin of the EBL}. Two major energy sources can contribute to the observed EBL:
nuclear and gravitational. The nuclear contribution consists of the energy released in stellar nucleosynthetic processes. 
This energy is radiated predominantly at UV$-$visual wavelengths and either redshifted or absorbed and
reradiated by dust into the infrared ($\lambda \gtrsim 1\ \mu$m) wavelength region. 
Gravitational potential energy dominates the energy released by brown dwarfs, accreting black holes, and gravitationally
collapsing systems, and may make a significant contribution to the EBL (see Bond, Carr, \& Hogan 1986, 1991, hereafter BCH86,
BCH91; Carr 1992, and references therein). Black holes in active galactic nuclei (AGNs) may contribute a significant fraction
of the EBL in the mid-IR ($\approx$ 10$-$50 \mic) wavelength region (Granato, Franceschini,
\& Danese 1996), and AGNs may be contributing to the IR energy released in starbursts as well. However, the magnitude of the AGN contribution 
to the diffuse IR background is highly
uncertain, since it depends on the amount of dust and the geometry of the torus around the active nucleus.
A third possible source of energy is that released by decaying relic particles (BCH86, 91). Since the infrared EBL should contain a
significant fraction of {\it all} the energy released in the universe since the recombination epoch, its measurement can constrain the
relative contribution of the various energy sources. For example, as discussed in Paper I (see also Figure 9 in this paper), 
the intensity predicted for some
non$-$nuclear energy sources (e.g. BCH86, BCH91)
 falls above the DIRBE upper limits, indicating that these models are not viable without significant
modifications. 

(2) {\it The evolution of galaxies}.
The intensity and spectrum of the EBL contains information on the evolution of the number
density, luminosity, and spectral energy distribution (SED) of galaxies over the history of the
universe. The number density of galaxies is expected to evolve as galaxies merge and undergo a
starbursting activity, or fade out of view. Deep galaxy counts using data obtained by the {\it
Infrared Astronomical Satellite} ({\it IRAS}) suggest either mild luminosity evolution, or
comoving density evolution with redshift up to
$\lesssim 0.01$ (Ashby et al. 1996). On theoretical grounds, the galaxy luminosity and SED are
functions of its star formation activity, metallicity, and dust content, all evolving quantities
that determine its far$-$IR appearance (e.g. Dwek
\& V\'arosi 1996; Dwek 1998). The {\it COBE} data can therefore be used to set constraints on
these evolutionary processes. 

\noindent
(3) {\it The star formation history of the universe}.
 The spectrally$-$integrated EBL at IR wavelengths should contain much of the energy  associated
with the production of elements throughout the history of the universe. The EBL spectrum depends
on the details of the cosmic star formation history, and the transport of the stellar energy
through the ambient dusty environment. {\it Hubble Space Telescope} ({\it HST}) observations of
the Hubble Deep Field (HDF) provided important new constraints on the star$-$formation rates
(SFR) at redshifts between $\sim$ 2 and 5 (Madau et al. 1996). These observations suggest that
the cosmic SFR at these redshifts falls significantly below those in the more local universe
(Lilly, Le F\'evre, Hammer, \& Crampton 1996), suggesting that the cosmic star formation activity
peaked at redshifts of about 1$-$1.5.  However, an important ongoing debate is whether the
UV and optically (UVO) derived SFR inferred from ground$-$based and {\it HST} observations severely
underestimates the actual star formation rate in the universe. Using {\it Infrared Space
Observatory} ({\it ISO}) observations of the HDF, Rowan$-$Robinson et al. (1997; hereafter RR97)
argued that the {\it HST} missed a significant fraction of the star formation  activity that takes
place in dust enshrouded galaxies at $z\gtrsim$ 2. The {\it COBE} data provide an important
constraint on the star formation history of the universe, and can therefore confirm (or rule out)
the presence of  dust$-$enshrouded galaxies or star$-$forming regions that may have led to an
underestimate of the cosmic SFR.

To address the origin of the EBL, we examine whether the intensity detected by the DIRBE and
FIRAS is consistent with that expected from the He$-$enrichment and metal formation in the
universe. Assuming a nuclear origin for the observed EBL and a cosmic star formation history, 
one can set a lower limit on the mass
fraction of baryonic matter that must have been processes into He and heavier elements in stars (\S 4). To examine whether
the {\it COBE} observations suggest any galaxy evolution, we develop a simple 
model to calculate the intensity and spectrum of the EBL (\S 5). Using a population synthesis model
 we calculate the spectral luminosity density produced by unattenuated starlight at each redshift. 
Then, for a given
 magnitude of visual extinction and a Galactic extinction law, we calculate the fraction of starlight that is 
absorbed by dust and converted into IR emission at each redshift. For simplicity, we assume that 
the extinction is constant as a function of redshift, and
 that the emitted IR spectrum is identical in shape to that produced by {\it IRAS} galaxies in the local universe.
The
spectrum of the EBL is then obtained by integrating the comoving spectral luminosity density of the attenuated starlight
 and the emitting dust over
redshift. By examining the extent to which the spectrum
calculated by this simple model deviates from the observed EBL, we can obtain clues to possible galactic 
evolutionary processes.  To
examine the constraints provided by the {\it COBE} on the cosmic star formation history, we
calculate the EBL spectrum for various star formation histories. The resulting EBL spectra 
(as well as those derived by other authors) are compared to the
observational constraints in \S 6.  The results of the paper are briefly summarized in \S 7. A value of 
$H_0$ = 50 km s$^{-1}$ Mpc$^{-1}$ for the
Hubble constant, and a flat universe ($\Omega_0=1$), with a zero cosmological constant
($\Omega_{\Lambda}$ = 0) is adopted throughout this paper.

\section{POSSIBLE ISOTROPIC LOCAL CONTRIBUTIONS TO THE 140 AND 240 $\mu$m RESIDUAL EMISSION}

Before studying the cosmological implications of the DIRBE results, we must examine if local sources
in the solar system or in the Galaxy can contribute a significant fraction of the observed residual
emission. 
Given a $\nu^n$ emissivity law, the temperature of a radiating particle is determined by the emissivity index,
$n$, and by the 140 and 240 $\mu$m intensities attributed to these sources.
Further constraints on the
temperature of the source are
 placed by the DIRBE 1.25 $-$ 100 $\mu$m upper limits (Paper I), and the FIRAS dark sky limits of $\sim$ 12.2 \hbox {and} 1.3
nW m$^{-2}$ sr$^{-1}$ at 340 and 650 $\mu$m, respectively (Shafer et al. 1997).

Figure 1 depicts the spectra of various hypothetical thermal sources that can contribute to the 140 and 240 $\mu$m residual
emission, for blackbody emitters characterized by an $n=0$ emissivity law (Fig 1a), and for dust particles characterized by an $n=2$
emissivity law (Fig 1b). The highest temperatures consistent with the DIRBE upper limits is 100 K for blackbody particles,
and 40 K for $n=2$ dust particles. Blackbodies with temperatures lower than 18 K will violate the FIRAS dark sky limits.
This temperature is slightly above the value of $\sim$ 16 K attained by such particles if they are only heated by the local ISRF. The first firm
conclusion that we can draw from these figures is that
 neither stars, stellar remnants, nor brown dwarfs can possibly contribute to the emission at these
wavelengths at the level required to account for a significant fraction of the residual emission.
To further characterize the properties of the potential foreground emission sources, we need to specify their physical 
location,i.e., whether they reside in the interstellar
medium (ISM) or in the solar system.

\subsection{A Solar System Isotropic Emission Component}

\subsubsection{The Required Characteristics of a Radiating Cloud of 
                Solar System Dust Particles}
To give rise to an isotropic emission component, any emitting sources within the solar system must have a spherically
symmetric distribution, either centered on the Earth, or centered on the Sun and located at a sufficiently large distance to
appear isotropic when viewed from the Earth. 
The measured DIRBE intensities were independent of zenith distance, ruling out an Earth$-$centered cloud of particles. 
Any local emission component must therefore be Sun$-$centered, a conclusion supported by the limits on the source
temperature and distance discussed below. We will also assume that such a cloud was assembled during, or shortly after, the formation of the
solar system (about
$\tau_{\odot}$ = 4.5 billion years ago), since it is unlikely that collisions could have produced a spherical distribution of
particles after most of the solar system material settled into a disk. 

Given an emissivity law and the particle temperature $T$, the cloud's
distance from the Sun, $d$, can be calculated from the energy balance equation:

\begin{equation}
\pi a^2 (1-A)\ \left[ {L_{\odot}\over 4 \pi d^2} + {c\over 4 \pi}\ U_{ISRF}\right ] = 4\pi a^2 \sigma T^4\ C(T)
\end{equation}

\noindent
where $a$ is the radius of the emitting particles, $A$ their effective albedo, $C(T)$ is the Planck-averaged value of their
emissivity, and $U_{ISRF}$ is the energy density of the local interstellar radiation field (ISRF). Assuming an albedo of 0.5, large
particles will attain temperatures of about 235~K at a distance of 1~AU from the Sun, whereas particles with optical
properties similar to interstellar dust particles (Draine \& Lee 1984) will have typical temperatures of $\sim$ 350~K. The
distance $d$ at which a particle attains a temperature $T$ can then be written as:

\begin{eqnarray}
d (AU) & = &  \left( {T\over 235}\right)^{-2}\ \ \ \hbox{for blackbody particles (}n \hbox{=0)},\ \ \ \nl   \nonumber 
  & = & \left({T\over 350}\right)^{-3}\left[1-\left({16\over T}\right)^6\right]^{-{1\over2}}\ \ \ \hbox{for dust particles
particles (}n \hbox{=2)}.\
\
\
\end{eqnarray}

\noindent
At any distance from the Sun, the particles will always be heated by the ISRF. This heating source is less important for
blackbodies, but gives dust particles a minimum temperature of about 16~K. Given the heliocentric distance of the cloud, its 140 or 240
$\mu$m brightness is given by

\begin{equation}
M_c = {4\pi d^2 I(\lambda) \over \kappa(\lambda) B_{\lambda}(T)},
\end{equation}

\noindent
where $\kappa(\lambda)$ is the mass absorption coefficient of the particles at wavelength $\lambda$. For 
blackbodies, $\kappa$ is wavelength independent and given by $3/(4\rho a)$, where $\rho
\approx$ 1 - 3 g cm$^{-3}$ is the mass density of the radiating particle. Small dust particles will be characterized by a
$\nu^2$ emissivity law and a value of $\kappa = 7.2$ cm$^2$ g$^{-1}$ at $\lambda$ = 240 $\mu$m (Draine \& Lee 1984).

The resulting cloud heliocentric distances (in AU) and masses are indicated in square brackets in Figures 1a and 1b. For blackbodies,
the masses were calculated for particles with 1 cm radius. Cloud masses scale linearly with particle 
radius, and also depend on the particle albedo and mass density. For $n$=2 dust particles, cloud masses only depend on the 
mass absorption coefficient and albedo. The figures show that cloud distances range from $\sim$ 5 to 170 AU for
$n$ = 0, ($\gtrsim$ 700 AU for $n$ = 2), and cloud masses range from $\sim 10^{22}a(cm)$ to $10^{27}a(cm)$ g
for $n$ = 0, ($\sim 10^{25}$ to
$10^{30}$ g for $n$ = 2). 
Depending on their origin, cloud particles could have a higher albedo and a lower mass density than assumed, as has been suggested
for Kuiper Belt particles (Teplitz et al. 1998). At a fixed temperature, namely that required to
produce the observed DIRBE detections (see Fig. 1), a lower albedo of $A \sim$ 0.05 will place the dust cloud
 at a larger distance, {\it increasing} the cloud mass by about a factor of 2 above that given in the figure.
On the other hand, a significantly lower mass density of $\sim$ 0.5 $-$ 1 g cm$^{-3}$, characteristic of
 icy particles, will {\it decrease} the cloud mass by a factor of $\sim$ 3$-$6. These numbers
 should serve as a guide for the uncertainties in the estimated cloud masses.  
For comparison, distances and masses of selected {\it known} solar system components are
given in Table 1 (Leinert 1996).
The limits on the source temperature and distance provide an additional argument that rules out the possibility that a
significant part of the DIRBE residual emission originates from any (T$\sim$~300~K) Earth$-$centered distribution of particles.

Any cloud of particles in the solar system will be subject to various forces and disruptive processes: radiation pressure
and solar gravity; thermal and kinetic sputtering, grain-grain collisions, and collisional drag
as the solar system moves through the interstellar medium; and gravitational
 perturbations from passing stars. Each one of these processes will tend to
erode or disrupt the cloud of particles. In Appendix $A$ we examine the lifetime of the cloud
against the various interactions with the sun, the ISM, and passing
 stars. 

\subsubsection{The Stability of a Cloud of Solar System Dust Particles}
Figure 2 shows the \{$a$(cm), $d$(AU)\} parameter space that must be occupied by any interplanetary material that could 
significantly contribute to the DIRBE residual 140 and 240 $\mu$m
emission. Shaded areas represent regions that are either excluded by
the temperature limits of the cloud or where the material would be eroded 
due to the various
effects considered in Appendix A on time scales shorter than the lifetime of the solar system. The horizontal lines marked ``high$-$T
limit'' and ``low$-$T limit" indicate the heliocentric distances below and above which the particle
temperatures fall beyond the values allowed in Figure 1a (particles with an $n = 0$ emissivity law). Both limits are
shifted to larger values of $d$ if the cloud consists of dust particles with an $n = 2$ emissivity law
instead. However, the figure shows that a cloud of dust particles with radii $\lesssim$ 0.1 cm is unstable against Poynting$-$Robertson (P$-$R)
drag and interactions with the ISM. Only a cloud consisting of objects larger than $\sim$ 1 cm
 located between $\sim$ 5 and 150 AU would be stable against local and ISM forces. A cloud with
a temperature of 30~K located at 60 AU could give rise to {\it all} of the residual isotropic emission observed
in the DIRBE bands provided its mass is
$\gtrsim\ 10^{26}$a(cm) g. The distance and temperature of the cloud are similar to those expected for the Kuiper Belt.
However, the Kuiper Belt is a disk or ring, and cannot produce an isotropic emission component. Solar system components that
can produce an isotropic foreground, such as the Inner and Outer Oort Clouds, are too distant and their emission is too faint
to contribute significantly to the DIRBE 140 or 240 $\mu$m emission. Table 1 is adapted from Leinert (1996) and shows the possible 240
$\mu$m contributions from known solar system components. The hypothetical cloud considered above is the last entry in the
table, marked ``Hypothetical Cloud''.

Such Hypothetical Clouds may have been observationally detected around main sequence stars by the {\it IRAS} satellite. From
detailed statistical analysis of the {\it IRAS} survey data of main sequence stars located within
$\sim$ 25 pc of the Sun, Aumann (1988) suggested that the presence of cool shells around A, F, and G stars is the
rule, rather than the exception. Further analysis (Aumann \& Good 1990) suggested that typical cloud temperatures
 are 20 $-$ 38~K, and cloud radii are
 100 $-$ 150 AU, not unlike the properties of the Hypothetical Cloud required to produce the residual
isotropic emission. The presence of these shells was inferred from the 60 $\mu$m excess in the colors of these
stars, and {\it IRAS} lacked the spatial resolution to determine the geometry of the cloud. In the following we show that a
spherical cloud at these distances, will be disrupted by collisions, and probably settle into a disk. The {\it IRAS} data are
therefore probably indicative of cool disks, rather than spherical shells, around these stars. This has been directly 
confirmed in some cases (e.g. $\beta$ Pic).  

Adopting cm$-$sized objects as the cloud constituent particles, it is easy to show that the
Hypothetical Cloud lifetime against internal collisions is significantly smaller than the age of the solar system. The
collisional lifetime is roughly given by $(n\ \Delta v\ \sigma)^{-1}$, where $n$ is the number density of cloud particles,
$\Delta v$ their relative velocities or internal velocity dispersions, and $\sigma$ their geometrical cross section. The
collisional lifetime is therefore proportional to the optical pathlength ${\it l} \equiv \tau/(n \sigma)$, and can
therefore be related to the residual intensity $I_{\nu}$ it is required to produce by:

\begin{equation}
\tau_{coll} = {B_{\nu}(T)\ l\over I_{\nu}\ \Delta v}
\end{equation}

\noindent
Adopting a cloud
distance, $d$, of
$60$ AU, we find that the orbital velocity $v$ is $\sim 4$ km s$^{-1}$. 
Perturbative forces due to other solar system objects will cause particle orbits to cross, so that $\Delta v \approx$ 4 km
s$^{-1}$. For $T = 30$ K, and a shell thickness ${\it l}$ equal to one$-$tenth of the clouds heliocentric distance, the
collisional lifetime is
$\sim 3\times10^6$ yr, considerably smaller than the lifetime of the solar system. This conclusion is not significantly
altered if we adopt different cloud distances or particle sizes within the range of distances allowed in Figure 2. The
internal collisions will grind the cloud particles into finer dust particles, causing them to either spiral into
the Sun, or settle into a disk$-$like configuration.

Having ruled out any {\it stable isotropic} cloud as a source of the residual 140 and 240 $\mu$m emission, we still need to
consider the possible existence of a cloud of particles that is continually replenished ({\it as is the main interplanetary
dust cloud}) as a viable source for the emission. Fig. 1b and Table 1 show that the Inner Oort
cloud is positioned at the correct distance and has a suitable isotropic geometry and mass for being a potential source of
the cloud. However, the transport of particles from this location to a distance of $\sim$ 100 AU by P$-$R drag will take over 10$^{12}$
yr, making any causal relation between the two clouds highly improbable. Other solar system components are even less likely
to be the source of the cloud since they are either farther away, or have too little mass. We therefore conclude that no
existing solar system component can possibly replenish the mass loss from a cloud sufficiently massive to produce significant
140 and 240 $\mu$m emission.

\subsection{A Galactic Isotropic Emission Component}

The temperature limits discussed above eliminate the possibility that a
significant portion of the 240 $\mu$m emission arises from warm and hot
Galactic sources such as white or brown dwarfs. The only significant
Galactic contribution to the 240 $\mu$m residual must come from
interstellar dust with a temperature $<$ 30 K, assuming an emissivity index of
$n=2$. By itself, this requirement can be easily met, since the 140 - 240 $\mu$m color
temperature of the residual emission is $\sim$16.5 K, which is quite close to
the temperature derived for dust associated with H I in the outer Galaxy
(Sodroski et al. 1997; Dwek et al. 1997). However, the isotropy of the residual emission
suggests that any significant contribution to the 240 $\mu$m residual
emission must come from a large (radius $R \gg 8.5$ kpc), roughly spherical,
shell of material. A smaller, or centrally condensed cloud would produce observable brightness variations
of the emission across the sky. 
It is very likely that stellar radiation pressure or correlated supernova explosions can expel dust into the halo of our
Galaxy (Ferrara, Ferrini, Franco, \& Barsella 1991). However, the expelled dust is likely to be patchy and associated with
Galactic fountains, chimneys, or clusters of OB stars, and therefore not likely to give rise to an isotropic emission
component. Furthermore, the required mass of expelled dust is too high to be of Galactic origin. If the entire 240
$\mu$m residual intensity is assumed to arise from a Galactic dust shell, then its mass is 
$M_{dust} \approx 10^5~R^2$  $M_{\odot}$ (see Eq. 4), where $R$ is the radius of the shell in kpc. Adopting a radius of 15
kpc, yields a total mass of
$\sim2\times10^7$ M$_{\sun}$ of dust and $\sim3\times10^9$ M$_{\sun}$ of gas
(assuming a dust-to-gas mass ratio of $Z_d$ = 0.007). This mass of gas is
nearly as large as the mass of the gas in the disk of the Galaxy (e.g., Sodroski et al
1997). This large gas mass suggests that it is very unlikely that the residual arises from an
unmodelled Galactic ISM component.

The dust required to produce an isotropic Galactic background component exhibits very similar characteristics to
 the dust tentatively detected by Zaritsky (1994) to be present in the halo of the galaxies NGC 2835 and NGC 3521.
At a distance of 15 kpc, the mass surface density of the hypothetical Galactic dust shell must be about 10$^{-6}$ g cm$^{-2}$, giving
 a visual optical depth of $\sim$ 0.05 for a Galactic extinction law. Such a halo of dust particles will produce
 a $B-I$ color excess of about 0.07, similar to the extinction observed by Zaritsky (1994) through the halo of the two galaxies at a distance
 of 60 kpc from their center. These observations are taken as preliminary evidence that galactic halos may be dusty even at those radii.
 However, the extinction observed by Zaritsky was along the major axis of these galaxies, and can therefore be
 only considered as evidence for the presence of dust in an extended disk. The mass of dust required to be in the disk and to produce the 
 observed color excess will therefore be significantly less than the mass of Galactic halo dust required to produce the intensity of the isotropic residual 
IR emission. The presence of a dusty disk component in our Galaxy, with characteristics similar to the two galaxies observed by Zaritsky, can 
therefore not be ruled out; however, 
such dust cannot produce an isotropic emission component.

We conclude from the considerations of this Section that the residual isotropic 140 and 240 $\mu$m radiation measured by the DIRBE 
and, by inference, the detections measured by the FIRAS as well,  
are most likely of extragalactic origin.

\section{LIMITS ON THE EBL AT 100 $\mu$m}

After the removal of foreground emission, the DIRBE
instrument detected a positive residual at 100
$\mu$m with a value of
$\nu I(\nu)$ = 21.9$\pm$6.1 nW m$^{-2}$ sr$^{-1}$ (Paper I).  However,
this residual emission is not isotropic, a strict requirement for the extragalactic background. Hauser et al. (1998)
therefore claimed a 95\% C. L. upper limit of $\nu I(\nu)\ <34$ nW m$^{-2}$ sr$^{-1}$ at this
wavelength. A lower limit of 3.9 nW m$^{-2}$ sr$^{-1}$ was derived from {\it IRAS} galaxy counts
(Hacking \& Soifer 1991). However, the positive detections at 140 and 240 $\mu$m can be used to
raise the lower limit on the IR intensity that must be present at 100 $\mu$m.
 
Infrared lines,  most notably the [C II] 158 $\mu$m line, will not contribute significantly to the
EBL detected in the DIRBE 140 and 240 $\mu$m bands. Because of the broad DIRBE spectral response, even the Galactic C$^+$
line does not contribute significantly to the energy received in the 140 $\mu$m band (Wright et al. 1991). Emission from
different redshifts will further reduce the relative line contribution to the in$-$band flux, since the line can
originate from only a limited range of redshifts whereas no such restriction applies to the continuum. The EBL detected in the
DIRBE 140 and 240 $\mu$m bands most likely consists largely of emission from dust heated by X$-$ray, UV, or optical photons,
regardless of their origin (whether stellar or non$-$thermal). The same physical process that produces the 140 and 240
$\mu$m continuum emission must contribute to a signal at 100
$\mu$m as well. The exact value of this contribution is, of course,
 model dependent. However, the steepest possible dropoff from 140 to 100 $\mu$m is the Wien exponential corresponding to the
coldest dust spectrum that is consistent with the 140 and 240 $\mu$m detections. The coldest dust spectrum is obtained by
fitting the DIRBE detections with a blackbody modified by a $\nu^2$ emissivity law. Any more
realistic spectrum, consisting of the contribution of emission from dust at different temperatures and
redshifts will give rise to a broader spectrum, and an increased 100 $\mu$m intensity.   

A fit to the nominal values of the DIRBE detections gives a dust temperature of about 15 K, and a lower limit of 
$\nu I(\nu)$ = 14  nW m$^{-2}$ sr$^{-1}$ at 100 $\mu$m. To assess the statistical significance of any lower limit, we
calculated the 100
$\mu$m intensity for all possible 
\{I(140 $\mu$m)$\pm\ 2\sigma$, I(240 $\mu$m)$\pm\ 2\sigma$\}, combinations, assuming partially (50\%) correlated errors. The
results show that $\nu I(\nu) >$ 5 nW m$^{-2}$ sr$^{-1}$ with a larger than 95\% probability. Hence, under the assumption that
most of the 140 and 240~$\mu$m residual emission is thermal emission from dust, we adopt the following
conservative lower and upper limits on the EBL at 100 $\mu$m:

\begin{equation}
5\ <\ \nu I(\nu)\ [\hbox {nW m$^{-2}$ sr$^{-1}$}]\ <\ 34\quad \quad  \hbox{at 100 $\mu$m}.
\end{equation}

\section{THE EXTRAGALACTIC BACKGROUND LIGHT FROM NUCLEOSYNTHESIS}

The total integrated Extragalactic Background Light detected by the DIRBE and FIRAS instruments
in the 140 $-$ 1000 $\mu$m wavelength interval is $\sim$ 16 nW m$^{-2}$ sr$^{-1}$. In
this Section we compare the integrated background with that expected from the stellar
production of helium and metals. Given a cosmic star formation rate, we calculate the
energy density in a comoving volume element as a function of redshift. The observed EBL consists
of the cumulative redshifted radiation from these volume elements.

The differential bolometric flux
$d{\cal F}$ received from a comoving volume element $dV_c$ located at redshift
$z$, with a comoving luminosity density 
$\epsilon(z)$ is given by (e.g., Kolb \& Turner 1990):

\begin{equation}
d{\cal F} = {\epsilon (z)\ dV_c(z) \over 4 \pi\ d^2_L(z)},
\end{equation}

\noindent
where $d_L(z)$ is the luminosity distance to the volume element, and 
\begin{equation}
{dV_c\over 4\pi d_L^2}\ =\left( {\delta \Omega\over 4\pi}\right)\left|{cdt\over dz}\right| {dz\over1+z}.
\end{equation}
The frequency$-$integrated intensity $I$ received from a distribution of sources within the solid angle $\delta \Omega$
 is given by the integral:

\begin{equation}
I = \left({c\over 4\pi}\right)\ \int_0^{z_*} \epsilon(z) \left|{dt\over dz}\right|{dz\over 1+z},
\end{equation}

\noindent
where $z_*$ represents the redshift when stars first turned on, and 
\begin{eqnarray}
\left|dt/dz\right|^{-1} & = & H_0 (1+z)\left[(1+z)^2(1+\Omega_0z)-z(2+z)\Omega_{\Lambda}\right]^{1/2} \nonumber
\nl
 & \equiv & \ H_0\ {\cal G}(\Omega_0,\ \Omega_{\Lambda},\ z)\ \ \ , 
\end{eqnarray}
\noindent
where $H_0$ is the Hubble constant, $\Omega_0\equiv
\rho_o/\rho_c$ is the present mass density of the universe normalized to the critical density  $\rho_c =
4.70\times10^{-30}\ h^2_{50}$ g cm$^{-3}$,
$\Omega_{\Lambda} \equiv \Lambda/3H^2_0$ is the dimensionless cosmological constant, and $h_{50}$ is the value of $H_0$ in
units of
$50$ km s$^{-1}$ Mpc$^{-1}$. 

It is useful to define a dimensionless radiative energy density parameter $\Omega_R$ as:
\begin{equation}
\Omega_{R} \equiv \left({4 \pi\over c}\right) I /\rho_c c^2,
\end{equation}
\noindent
where
\begin{equation}
\rho_cc^2 = 4.23\times10^{-9}\ h_{50}^2\ \ \ \hbox{erg\ cm}^{-3}   .
\end{equation}

\noindent
Numerically, the parameter $\Omega_{R}$ is related to the integrated EBL intensity $I$ as:
\begin{equation}
\Omega_{R} = 1.0\times10^{-7}\ I(\hbox{nW\ m}^{-2}\ \hbox{sr}^{-2})\ h_{50}^{-2}.\ \ 
\end{equation} 

To calculate $I$, we need to know the value of the luminosity density, $\epsilon$, as a function of redshift $z$, or
epoch
$t_z$ measured since the epoch $t_*$, when stars first turned on. Assuming that the total radiant background arises
from energy released by nucleosynthesis, the value of
$\epsilon$ is related to the cosmic star formation rate
$\psi(t)$ (in stars yr$^{-1}$ Mpc$^{-3}$) as

\begin{equation}
\epsilon (t_z) = \int_{t_*}^{t_z} dt\ \psi(t)\ \int_{M_{low}}^{M(t')}\ L(m,t') \phi(m) dm,
\end{equation}

\noindent
where $L(m,t')$ is the luminosity of a star of initial main$-$sequence mass $m$ at time
$t'=t_z-t$, $\phi(m)$ is the stellar mass spectrum, normalized to unity in the \{M$_{low}$,
M$_{up}$\} mass interval, and $M(t')$ is the initial main$-$sequence mass of a star with a lifetime $t'
= t_z - t$. Equation (14) was evaluated using the stellar evolutionary tracks of Bressan,
Fagotto, Bertelli, \& Chiosi (1993), and takes the delayed release of stellar energy
into account. Consequently, even if all stars were formed in an instantaneous burst, their energy
output will be spread out over their respective lifetimes.  

The dependence of the cosmic star formation rate on redshift can be inferred from the variation
of metal abundances in QSO absorption$-$line systems (e.g., Pettini, Smith, King, \& Hunstead
1997), from the consumption of H I in damped Ly$\alpha$ systems (Pei \& Fall 1995, hereafter PF95), or from
photometric studies of the UV output of galaxies at various redshifts (e.g., Madau et al. 1996,
and references therein). Applying the UV and blue dropout techniques to galaxies in the Hubble
Deep Field (HDF), Madau et al. (1996) derived star formation rates for $z = 2.75$ and
$4.0$. Figure 3 depicts several cosmic star formation rates as a function of redshift.
The UVO curve represents the UV$-$optically derived cosmic star formation rate compiled by 
Madau et al. (1996) as recently revised by Madau, Pozetti, \& Dickinson (1997, hereafter MPD97).
 The revised SFR is somewhat higher
that the previous estimate of Madau et al. (1996), peaking at a somewhat higher redshift of
$z\approx 1.4$. The PFI and PFC curves represent the cosmic SFR in, respectively, 
the infall and closed box cosmic chemical evolution models of PF95. Figure 1d in PF95 depicts the {\it net}
 mass consumption rate due to star formation, which needs to be divided by ($1-R$), where $R$ is the average fraction of the 
initial stellar mass that is returned back to the ISM over the stellar lifetime. In deriving the stellar birthrate
 we assumed this fraction is constant and equal to 0.30. 
For a Salpeter stellar mass spectrum, $\phi(m)
\sim\ m^{-2.35}$, and \{M$_{low}$, M$_{up}$\} = \{0.1 M$_{\odot}$, 120 M$_{\odot}$\}, the average stellar mass
is 0.35 M$_{\odot}$, so the stellar birthrate, $\psi$ (in yr$^{-1}$ Mpc$^{-3}$), is related to the mass consumption rate due 
to star formation, $\rho_*$, as: 
\begin{equation}
\psi (yr^{-1}\ Mpc^{-3}) = 2.9\times \rho_*(M_{\odot}\ yr^{-1}\  Mpc^{-3}). 
\end{equation}
\noindent

The star formation rate derived from UV and optical observations is, in principle, a lower limit on
the cosmic SFR. Ultraviolet and optical surveys systematically underestimate the star formation
rate, since a significant fraction of the UV$-$optical stellar output can be absorbed and
reradiated at IR wavelengths by dust. The magnitude of the discrepancy is, however, still
controversial. Rowan$-$Robinson et al. (1997) argued that ISO observations of the HDF  suggest
that the star formation rate remains constant at
$z\gtrsim 1.5$, instead of decreasing. In this scenario the UV observations of the HDF represent
only a fraction of
 of the actual star$-$formation activities at these high redshifts. 
The actual SFR may have been underestimated at lower redshifts as well.
Figure 4 depicts the cosmic luminosity density, calculated using eq. (14), as a function of
redshift for the various  cosmic star formation rates depicted in Figure 3. The various star formation scenarios
give somewhat different local luminosity densities which are summarized and compared to the
observations in Table 2.

Integration of the luminosity density over redshift (eq. 9) for $\Omega_0 = 1$ and
$\Omega_{\Lambda} = 0$ gives
\begin{eqnarray}
I \hbox {(nW m}^{-2}\ \hbox {sr}^{-1}\ \hbox {)}  & = & 30\ \  \quad \quad	\hbox {for UVO} \nonumber \\
    & = & 91\ \  \quad \quad	\hbox {for PFI} \\
    & = & 41\ \  \quad \quad	\hbox {for PFC} \nonumber
\end{eqnarray}
 
\noindent
The positive detection of the EBL by the {\it COBE} therefore accounts for about 20 to 50\% of
 the expected EBL associated with the energy release from nucleosynthesis, depending on the adopted cosmic SFR. 
The EBL was also recently detected at UV$-$optical wavelengths by Bernstein (1997, see also Bernstein,
Freedman \& Madore (1998). Furthermore, Pozzetti et al. (1998) set a lower limit on the EBL at
 $\lambda$ = 0.36, 0.45, 0.67, 0.81, and 2.2 $\mu$m by calculating the contribution of discrete objects.
The Pozzetti et al. (1998) lower limits suggest an EBL intensity $\gtrsim$ 12 nW m$^{-2}$ sr$^{-1}$ in the 0.36
 to 2.2 $\mu$m wavelength interval, which combined with the {\it COBE} detections gives
 an {\it observed} EBL intensity in the 0.36$-$2.2 and 140$-$5000 $\mu$m wavelength intervals in excess 
of $\sim$ 28 nW m$^{-2}$ sr$^{-1}$. This value is similar to the EBL intensity predicted by the UVO cosmic
SFR. The UVO star formation rate therefore leaves no room for any expected EBL emission in the 2.2 to 140 $\mu$m wavelength region,
 indicating that the UV$-$optically determined SFR underestimates the actual rate of star formation
 in the universe. 
  
Most of the energy radiated by stars is liberated during the transmutation of protons into helium
and heavier elements. The radiative energy density parameter,
$\Omega_R$, can therefore be related to
$\Omega_*$, the fraction of the total critical mass density that has been processed through stars
(BCH86, Peebles 1995). The value of $\Omega_*$ depends on details of the chemical evolution of a
comoving volume element in the universe (Pei \& Fall 1995). For simplicity we will assume that
all the elements were instantaneously formed at some epoch corresponding to some redshift
$z_e$. The intensity of the EBL consists of the energy released from the production of He that
was not
 further processed into heavier elements, and of the energy released by the production of 
elements heavier than He. It can be written as:
 \begin{equation}
 I  =  \left( {c\over 4\pi}\right) {\Omega_* \rho_c c^2\over 1 + z_e}\ \ (\eta_Y \Delta Y\ +\ \eta_Z Z_*) \quad . 
\end{equation}

\noindent
The parameter $\eta_Y = 0.0072$ is the energy conversion efficiency for the nuclear energy
generating reactions 4p$
\rightarrow\  ^4$He, $\eta_Z = 0.0078$ is that for the 12p$ \rightarrow\ ^{12}$C reaction that
leads to the production of metals, $Z_*$ is the current mass fraction of matter that was
converted into metals, and
$\Delta Y$ is  the net enrichment in the $^4$He mass fraction due to stellar processing. For $Z_*
= Z_{\odot}=0.02$, the solar metallicity, and
$\Delta Y$ = 0.04, which is the difference between the solar ($Y = 0.28$) and the primordial ($Y =
0.24$) $^4$He mass fraction, most of the contribution to the EBL is due to the net enrichment of
He in the universe, and

\begin{equation}
\Omega_* = 6.3\times10^{-3}h_{50}^{-2}\ \left({\Omega_R\over 28\times 10^{-7}}\right)\ (1+z_e).
\end{equation}

\noindent
where $\Omega_R$ is normalized to the {\it observed} radiative energy density parameter of the EBL.

To put this derived value of $\Omega_*$ in perspective, we note that a strict upper limit on
$\Omega_*$ is provided by the baryonic mass$-$density parameter
$\Omega_{BBN}$ derived from Big Bang nucleosynthesis (BBN) arguments:
\begin{equation}
\Omega_* h_{50}^2\ <\ \Omega_{BBN} h_{50}^2\  \approx\ 0.044\ -\ 0.15\ \ \ , 
\end{equation}
where we adopted a conservative range of values for $\Omega_{BBN} h_{50}^2$ (Kolb \& Turner 1990;
Steigman, Hata,
\& Felten 1997). $\Omega_{BBN}$ represents a strict upper limit on $\Omega_*$, since a fraction
of the baryonic matter is locked up in low mass stars that never cycled their nucleosynthetic
products back to the interstellar medium. An approximate lower limit on
$\Omega_*$ can be derived from the amount of luminous matter currently locked up in stars:
\begin{equation}
\Omega_*\ >\  \Omega_{LUM} \equiv {\epsilon_B} <M/L_B>/ \rho_c \approx 1.4\times10^{-3}\ h_{50}^{-1},
\end{equation}

\noindent
where the numerical value was obtained for a local blue luminosity density, $\epsilon_B =
7.9\times10^6\ h_{50}$ L$_{\odot}$ Mpc$^{-3}$ (Lilly et al. 1996), presented here in units of the
bolometric solar luminosity L$_{\odot} = 3.826\times 10^{33}$ erg s$^{-1}$. We also used a
conservative lower limit of
$<M/L_B> = 14$ M$_{\odot}$/L$_{\odot}$ (in bolometric L$_{\odot}$) derived from stellar
population models of Larson \& Tinsley (1978). So 
we find the consistent relationships 
\begin{equation}
\Omega_{LUM}\ \approx 1.4\times10^{-3}\ h_{50}^{-1} <\ \Omega_*\approx 0.013\ 
                 h_{50}^{-2}\ {1+z_e\over 2} <\ \Omega_{BBN}\ \lesssim 0.15\ h_{50}^{-2}\ \ ,
\end{equation}
where we have included in $\Omega_R$ the UV and near$-$IR contribution to the EBL.
The average redshift for metal production, weighted by the integrand of eq. (9) is $z_e = 1.0$,
giving $\Omega_*\approx 0.013\ h_{50}^{-2}$ for the observed intensity of the EBL.
 This value of $\Omega_*$ is consistent with the range of lower limits of (0.044$-$0.15) h$_{50}^{-2}$ required for
Big Bang nucleosynthesis. The UV to near$-$IR and the {\it COBE} 140$-$5000 $\mu$m wavelength detection therefore
imply that at least $\sim$ 10\%, and possibly as much as $\sim$ 30\% of the total baryonic mass density inferred from Big Bang
nucleosynthesis was processed in stars into heavier elements. 
In the following we examine the consistency of the {\it COBE} background
measurements with the expected spectrum of the EBL.

\section{SPECTRAL ENERGY DISTRIBUTION OF THE EXTRAGALACTIC
         BACKGROUND LIGHT}

\subsection{Model Description}
Numerous models have been constructed by various authors to predict the intensity and spectrum of
the EBL using observational constraints and theoretical models for the evolution of galaxies.
Pioneering work in this field was conducted by Partridge \& Peebles (1967), Harwit (1970), BCH86,
BCH91, and Negroponte (1986; and references therein).  A more complete set of references to more
recent work can be found in various papers in the proceedings of the conference ``Unveiling
the Cosmic Infrared Background" (Dwek 1996).

The specific intensity
$I(\nu_0)$ of the EBL at the observed frequency $\nu_0$ is obtained by integrating the 
spectral luminosity density, $\epsilon(\nu, z)$, from the comoving volume elements at $z$, 
over redshifts:

\begin{equation}
I(\nu_0) = \left({c\over 4\pi}\right)\ \int_0^{z_*} \epsilon(\nu,z) \left|{dt\over dz}\right|dz
 \quad , 
\end{equation}

\noindent
where $\nu =\nu_0 (1+z)$ is the frequency in the rest frame of the comoving volume element.

At each redshift, the spectral luminosity density can be written as the following sum:

\begin{equation}
\epsilon(\nu,z) = \epsilon_s(\nu,z) + \epsilon_d(\nu,z) \quad ,
\end{equation}

\noindent
where $\epsilon_s(\nu,z)$ and $\epsilon_d(\nu,z)$ are, respectively, the attenuated 
starlight and the dust contribution
 to the spectral luminosity density. The total, unattenuated, spectral luminosity density from
starlight, $\epsilon(\nu,z)$, can be calculated from population synthesis models.
 Given the evolution of the cosmic SFR as a function of time, the stellar mass spectrum, stellar
evolutionary tracks and nucleosynthesis yields, and stellar atmosphere models as a function of
metallicity, the spectral luminosity density can be uniquely determined as a function of time
  (e.g. Charlot, Worthey \& Bressan 1996).
The redshift dependence of
 $\epsilon(\nu,z)$ is, to first order, independent of the cosmological parameters H$_0$,
$\Omega_0$,
 and $\Omega_{\Lambda}$, if the cosmic SFR is observationally determined or constrained as a
function
 of redshift. However, a second order dependence of the spectral luminosity density on 
the cosmological parameters enters into the model through the finite stellar lifetimes in the 
population synthesis models.

Significantly more complicated is the determination of the fraction of the stellar spectral
energy distribution that is absorbed by dust and reradiated at IR wavelengths. Many different 
approaches have been applied to the problem. Lonsdale (1995, 1996) grouped the
various methods into two major categories: Backward and Forward Evolution models. Cosmic Chemical
Evolution models (Pei \& Fall 1995) can be regarded as a third, distinct, category.
\underline {Backward Evolution} models adopt the currently observed galaxy luminosity function
and spectral energy distribution (see \S 5.2 below), and translate them backwards in time. 
Notable models in this
category were constructed by Hacking \& Soifer (1991), Beichman \& Helou (1991), and more
recently by Malkan \& Stecker (1997).
\underline {Forward Evolution} models evolve various galactic systems, such as elliptical,
spiral, and starburst galaxies, from an initial redshift
$z_*$ forward in time using
 population synthesis and chemical evolution models (e.g. Rocca$-$Volmerange \& Fioc 1996,
Franceschini et al. 1996, 1997; Guiderdoni et al. 1997). In these models, the fraction of absorbed
 starlight is determined by simple radiative transfer models, and the reradiated IR emission 
is taken as a combination of spectra representing dusty H~II, H~I, and molecular cloud regions.
 These models have been quite successful in reproducing the UV to far$-$IR spectra of various types of
 galaxies.  
\underline {Cosmic Chemical Evolution} models adopt a more global approach than Forward Evolution
models, by generalizing standard Galactic chemical evolution models (e.g. Tinsley 1980) to the
universe as a whole. The cosmic SFR in these models is constrained by the diminution of H~I column
 density as a function of redshift. Recently, Fall, Charlot, \& Pei (1996) used this approach in
conjunction with population synthesis, and dust and chemical evolution models to predict the EBL
at UV to far$-$IR wavelengths. Unique to these models is the fact that the gas opacity (and hence
the thermal
 dust emission) at each redshift is constrained by the observed obscuration of quasars.
   
In this paper we calculate the EBL spectrum using a combination of all three approaches. Given a cosmic star formation rate, 
the unattenuated stellar spectral luminosity density, $\epsilon(\nu,z)$, at each redshift is calculated
 using detailed population
 synthesis models. For these calculations, we adopt the stellar evolutionary tracks of Bressan et al. (1993), Kurucz
 stellar atmosphere models for a solar metallicity composition, and a Salpeter stellar mass spectrum
 in the 0.1 to 120~M$_{\odot}$ mass interval.
The attenuated stellar spectral luminosity density at each redshift is given by
\begin{equation}
\epsilon_s(\nu,z) = [1-P_{abs}(\nu)]\times \epsilon(\nu,z)\ \ \ ,
\end{equation}
\noindent
where $P_{abs}(\nu)$ is the probability that a photon of frequency $\nu$ is locally absorbed by the ambient dust.
In this model we take $P_{abs}(\nu) = \exp[-\tau_{abs}(\nu)]$, independent of redshift, where $\tau_{abs}$ is 
the dust absorption opacity, calculated for an average Galactic interstellar extinction law normalized to some value in the V band, 
 with a wavelength independent albedo of 0.5 (e.g., Dwek 1997).
Note that in spite of the fact that P$_{abs}(\nu)$ is independent of redshift, the fraction of starlight that is converted to 
IR emission is a redshift dependent quantity, since the spectrum of the stellar luminosity density evolves with time. 
We further assume that the spectrum of the emerging IR emission is given by the luminosity function$-$averaged spectrum of
 the {\it IRAS} galaxies (see \S5.2 and Fig. 5). 
The spectral luminosity density of the dust at $z$ can then be expressed in terms of its {\it local} spectral luminosity 
density, $\epsilon_d(\nu,0)$, as:

\begin{equation}
\epsilon_d(\nu,z) = \epsilon_d(\nu,0) \ { \int P_{abs}(\nu) \epsilon(\nu,z)d\nu \over 
                                     \int P_{abs}(\nu) \epsilon(\nu,0)d\nu}  
\end{equation}

The adopted prescription for calculating the IR background is
admittedly simplified. The cosmic SFR represents a statistical average
over many galactic systems, each possesing a complex star formation
history that is best studied with nonlinear multipopulation models
(see review by Shore \& Ferrini (1995). A more realistic approach to calculating galactic
spectra should include the effects of the
evolution of the dust abundance and composition, as well as the
clumpiness and fractal nature of the interstellar medium on the photon escape
probability (Fall, Charlot, \& Pei 1996; Dwek \& V\'arosi 1996; Dwek 1998; Witt \& Gordon 1996).

\subsection{The Local UV to Far$-$IR Spectral Luminosity Distribution}
The values of $\epsilon_s(\nu,z)$ and $\epsilon_d(\nu,z)$ at $z$=0 can be determined from
observations
 of the local luminosity density in the various visual, near$-$ and far$-$IR bands.  Using the 
Canada$-$France Redshift Survey, Lilly et al. (1996) estimated local luminosity densities of
 $\epsilon(\nu,0) = 3.0\times10^{25}, 2.0\times10^{26}$, and
$8.1\times10^{26}$ erg s$^{-1}$ Hz$^{-1}$ Mpc$^{-3}$ at $\lambda$ = 0.28, 0.44, and 1.0 $\mu$m,
respectively, for $h_{50}=1$. At 2.2
$\mu$m, Gardner et al. (1997) give a value of $1.1\times10^{27}$ erg s$^{-1}$ Hz$^{-1}$
Mpc$^{-3}$. The stellar spectral luminosity density can be well represented by the spectrum of an
average spiral or elliptical galaxy (Schmitt et al. 1997). Normalizing the spectrum of the average spiral
galaxy to the local spectral luminosity density yields an integrated luminosity density in the 0.1 to 10 $\mu$m
 wavelength interval of
$(1.30\pm 0.7)\times10^8$ L$_{\odot}$ Mpc$^{-3}$, the error reflecting the 1$\sigma$ uncertainties in
 the determination of the B-band luminosity density, including the value derived by MPD97 from the 
local B-band luminosity function determined by Ellis et al. (1996).
 Figure 5 depicts the data of Lilly et al. (1996) and Gardner et al. (1997), and the SED of an
average spiral galaxy (Schmitt et al. 1997, Table 6) normalized to fit the observed spectral
luminosity density
 at $\lambda \lesssim$ 3 $\mu$m. 

At wavelengths above $\sim 5-10\ \mu$m, the spectral luminosity density is dominated by thermal
emission from dust, and
 is represented by that of {\it IRAS} galaxies. In general, any galactic dust spectrum can be
represented by a combination of three emission components: (1) a hot dust component, as one would
expect from dust in H~II regions; (2) a warm dust component, as one would expect from dust in the
diffuse H~I gas and heated by the general interstellar radiation field; and (3) a cold dust
component, arising from dust in molecular clouds.

{\it IRAS} observations of IR bright galaxies show a systematic variation of IR colors with luminosity, characterized by decreasing
$I$(12$\mu$m)/$I$(25$\mu$m), and increasing $I$(60$\mu$m)/$I$(100$\mu$m) flux ratios, with increasing IR
luminosity (Soifer \& Neugebauer 1991). This trend can be used to construct luminosity$-$dependent spectral templates of
the {\it IRAS} galaxies using a linear combination of the three spectral dust components. In practice, the {\it IRAS}
color$-$color trend can be reproduced with just the H~II and H~I dust components. Figure 6 shows the H II and H I spectral
dust components, arbitrarily normalized. The H II spectrum represents that of
typical Galactic H II regions such as Orion (Wall et al. 1996). The H I spectrum represents the model fit to the
DIRBE observations of the diffuse ISM, consisting of the emission from a mixture of polycyclic aromatic hydrocarbons (PAHs),
graphite, and silicate grains, stochastically heated by the ambient radiation field (Dwek et al. 1997). Also shown in the
figure is the spectrum of Arp 220, an interacting ultraluminous IR
galaxy (ULIRG) obtained by Klaas et al. 1996 and Fischer et al. 1997 using the {\it Infrared Space Observatory} ({\it ISO}) satellite. 
Arp 220 is somewhat hotter than other interacting systems (Klaas et
al. 1996), so the H~II spectrum adopted in this paper is representative of a  prototypical ULIRG. Table 3 shows the relative
contribution of the H~II and H~I components to the IR spectra of various {\it IRAS} galaxies. The mean observed colors of the {\it IRAS}
galaxies as a function of their luminosity were taken from Table 5 of Soifer \& Neugebauer (1991). Galaxies with L$_{IR} =
10^{13}$ L$_{\odot}$ were represented by a pure H II spectrum. Figure 7 depicts the template spectra of these various
galaxies. The spectra of the more luminous galaxies peak at shorter wavelengths, an effect that is the result of the increased
contribution to their spectrum of the H~II component, relative to that of the H~I. 

An average local IR spectrum was derived by averaging the
individual galactic spectral templates over the {\it IRAS} galaxy luminosity function $\Phi(L)$. This luminosity
function (LF) was originally derived by Soifer et al. (1987). Here we use the fit of Beichman \&
Helou (1991), adjusted for a value of H$_0$ = 50 km s$^{-1}$
Mpc$^{-1}$:

\begin{eqnarray}
\Phi(L)\ dL & = & \Phi_* \left({L\over L_0}\right)^{\beta} \left({dL\over L_0}\right)\ \  for\ \ L \leq L_0 \nonumber \nl
               & = & \Phi_* \left({L\over L_0}\right)^{\gamma}\left({dL\over L_0}\right)\ \  for\ \  L \geq L_0
\end{eqnarray}

\noindent
where $L$ represents the IR luminosity of the galaxies, $L_0 = 5.65\times10^{10}$ L$_{\odot}$,
$\Phi_*=3.0\times10^{-4}$ Mpc$^{-3}$, $\beta = -1.65$, and $\gamma = -3.31$. An IR
luminosity range of $3.2\times10^8$ to $1\times10^{13}$ L$_{\odot}$ was adopted in all calculations.

The local spectral luminosity density from dust and stars as a function of wavelength is shown in
Figure 5. The spectrally integrated luminosity density of {\it IRAS} galaxies is $0.53\times10^8$
L$_{\odot}$ Mpc$^{-3}$, which, together with the stellar UV to near$-$IR local luminosity density
of $1.3\times10^8$ L$_{\odot}$ Mpc$^{-3}$, gives a total observed luminosity density
 of $\epsilon(\nu,0)\approx (1.8\pm 0.7)\times10^8$ L$_{\odot}$ Mpc$^{-3}$. We assumed here that the uncertainties in the 
local luminosity density are dominated by those in the stellar component. 
Table 2 compares the current luminosity density to that
 predicted by the various cosmic star formation scenarios discussed in \S 4. The table shows the PF95 infall model predicts
 a local luminosity density of $\sim 4\times10^8$ L$_{\odot}$ Mpc$^{-3}$, significantly larger than the observed value. The model can, however, 
be modified to fit the observational constraint if, for example, a fraction of the star formation activity 
at some epoch results in the production of massive stars that release all of their radiative energy instantaneously
 back into space. All other star formation scenarios reproduce the local luminosity density within the uncertainty of
 the observations.

The observed $\epsilon_d(\nu,0)/\epsilon(\nu,0)$ ratio is 0.30, suggesting that locally, about 30\% of the stellar
 light is processed by dust into thermal IR emission. The attenuated stellar spectral luminosity density, 
$\epsilon_s(\nu,0)$, can be derived by propagating the spectral luminosity density
 of the sources, $\epsilon(\nu,0)$, through a foreground screen characterized by a Galactic extinction law
 with a visual optical depth of $\tau_{abs}(V)\approx 0.27$.

\section{ASTROPHYSICAL IMPLICATIONS}

\subsection{Summary of EBL Detections and Observational Constraints}
 
Figure 8 depicts the current detection and observational constraints on the EBL at UV to far$-$IR
wavelengths.  The figure shows the DIRBE 140 and 240 $\mu$m detections plotted with 2$\sigma$
error bars (Paper I), the 100
$\mu$m lower limit estimated in \S 3, and the FIRAS 125 to 5000
$\mu$m detection (Fixsen et al. 1998, plotted here only between 200 and 1000 $\mu$m). Figure 8 also shows
 the DIRBE 2$\sigma$ upper limits in the 1.25$-$100
$\mu$m wavelength region, where foreground emissions from interplanetary dust and the Galaxy were
the main obstacles for the detection of any extragalactic signal. Upper limits on the intensity of
the EBL can also be derived from a comparison of the spatial fluctuations in the
DIRBE maps with those expected from spatial fluctuations of galaxy clusters (Kashlinsky, Mather,
Odenwald,
\& Hauser 1996). In the 10
$-$ 100 $\mu$m wavelength region, the fluctuation analysis suggests an upper limit of 10$-$15 nW
m$^{-2}$ sr$^{-1}$ (Kashlinsky, Mather \& Odenwald 1996; hereafter KMO96). 

TeV $\gamma-$ray observations of active galaxies can, in principle, provide additional
constraints on the intensity of the EBL at IR wavelengths (e.g., Stecker \& De Jager 1993). TeV
$\gamma-$rays interact primarily with 5$-$40 $\mu$m photons from the EBL by $e^+e^-$ pair
production. Any evidence for absorption in the spectra of the various TeV
$\gamma-$ray sources could therefore be used to infer the energy density of the EBL in this
wavelength region. In practice, these calculations require knowledge of the intrinsic
$\gamma-$ray spectrum of the source, and the shape of the EBL in the $\sim$~5$-$40 $\mu$m
wavelength regime (Dwek \& Slavin 1994). Currently, there is no evidence for any intergalactic
absorption in the spectrum of the three TeV sources detected to date (Krennrich et al. 1997,
Catanese et al. 1997). Claimed detections of, or upper limits on the EBL from TeV $\gamma-$ray
observations should therefore be regarded as preliminary at present. Nevertheless, we show
in Fig. (7) the 15
$-$ 40 $\mu$m upper limits derived from the $\gamma-$ray spectrum of Mrk 421 (Dwek \& Slavin
1994), and the 5 $-$ 25 
$\mu$m  upper limits derived from the observations of Mrk 501 (Stanev \& Franceschini 1997).
While these limits are well below the direct observational limits of the DIRBE (Paper I), they
are no stricter than those obtained by KMO96, and considering their uncertainty, provide soft
 constraints on currently popular models in this wavelength interval.

Figure 8 also shows the UV and optical lower limits derived from {\it HST} observations
of the HDF (Pozzetti et al. 1998), the near$-$IR lower limits from galaxy counts (Gardner et al.
1997), the 25, 60, and 100 $\mu$m lower limits from {\it IRAS} galaxy counts (Hacking \& Soifer 1991,
Gregorich et al. 1995), and the recent optical detections at 0.3, 0.55 and 0.8 $\mu$m by
Bernstein (1997), and Bernstein, Freedman, \& Madore (1998).

\subsection{Comparison of Observations With Model Calculations}

Figure 8 compares the EBL detections and constraints with various model predictions in the 0.1 to
1000 $\mu$m wavelength regime. The thick solid line represents the spectrum of the EBL predicted by 
our hybrid Forward$-$Backward$-$Cosmic Chemical Evolution model described in \S5.1, using the UVO
 cosmic SFR (hereafter referred to as the $UVO$ model). The discontinuities in the flux at 7.7, 8.6, and 11.3
$\mu$m results from the cumulative contribution of the respective PAH emission lines to the EBL.  
Also shown in the figure are the Backwards Evolution models of Beichman \& Helou (1991; $BH$) and
Malkan
\& Stecker (1997; $MS$); the Forward Evolution model of Franceschini et al. (1997; $FR$) and
Guiderdoni et al. (1997; $GD$); and the cosmic chemical evolution infall model of Fall, Charlot, \& Pei
(1996; $FCP$).
The curve labeled ``$S$" represents a ``dust$-$free" model, calculated with the UVO star formation rate 
in which galactic starlight is
allowed to escape into space unimpeded by dust. The curve illustrates the obvious, namely that in
order to have {\it any} significant far$-$IR emission, a fraction of the starlight has to be
reprocessed by dust.

Postponing the discussion of the results of the $UVO$ model for the moment, in a broad sense, all other models depicted 
in the figure, with the exception of the $BH$ and $FCP$ models, fall within 3$\sigma$ of the
DIRBE 140 $\mu$m detection, and all models fall within 3$\sigma$ of the 240 $\mu$m detection.
Most models are also consistent with the FIRAS detection of Fixsen et al. (1998), except for
models $GD$, and $BH$, which predict a somewhat higher flux at wavelengths
$\gtrsim$~500~
$\mu$m. The various models differ most in their predicted 5 $-$ 60
$\mu$m intensities. The $FCP$ model falls especially short in this wavelength region, since
(admittedly) the authors did not attempt to model the dust emission at these wavelengths.

Very obvious in the figure is the failure of the UV$-$optically determined star formation rate
 to produce the observed EBL at far$-$IR wavelengths. 
Observationaly this should not be too surprising since the most intense star forming regions frequently manifest 
themselves primarily 
through their infrared, rather than their UV or optical, emission, as recently illustrated by ISO observations 
of the Antennae galaxies (Mirabel et al. 1998).
Furthermore, we have already shown that the 
total integrated EBL intensity predicted by the UVO cosmic SFR was very close to the detected EBL at 
0.36$-$2.2 $\mu$m and 140$-$5000 $\mu$m, leaving no room for any EBL intensity in the $\sim 2 - 140
\ \mu$m 
region  of the spectrum. Since our template dust spectrum includes significant emission in this
 spectral wavelength region, the far$-$IR intensity predicted by the $UVO$ model must 
fall significantly below the observations. The calculated far$-$IR intensity cannot 
be increased by simply 
increasing the fraction of starlight that is reradiated in the model as thermal IR emission. This will require
an increase in the extinction, causing the  UV$-$optical part of the EBL to fall below the lower
limits of Pozzetti et al. (1997). It seems clear that the excess IR
emission
 needed to account for the detected EBL requires a star formation activity that escaped detection
 at UV and optical wavelengths.

\subsection{Are There UV$-$Optically ``Hidden" Galaxies or Star Forming Regions in the Universe?}

The existence of optically obscured star forming regions is not only suggested by the discrepancy of the $UVO$ model
 EBL spectrum and the observations. Figure 8 shows that {\it all} model predictions are 2$\sigma$ below the DIRBE 
EBL measurement at 140 $\mu$m, which is consistent with the somewhat noisier FIRAS determination at this
wavelength. A population of optically$-$hidden, dust enshrouded 
galaxies or star forming regions, may therefore be needed to account for the observed infrared component of the EBL in all models.

We therefore considered various additional star formation histories. 
The cosmic SFR predicted
 by the infall or closed box models of PF95 (labeled PFI and PFC, respectively, in Figure 3) offer
 alternatives to the $UVO$ model, since they a priori predict higher values for the 
integrated EBL intensity (see \S4 and Table 2). 
Since the integrated EBL of the $PFI$ model is significantly larger than that predicted by the $UVO$ model, 
it can produce the observed infrared EBL with less extinction than the $UVO$ model, whereas the $PFC$ model requires 
a somewhat larger extinction to reproduce the observations. 
We therefore calculated the EBL spectrum using the $PFI$ and $PFC$ star formation rates
 with $\tau_{abs}(V)$ values of 0.20 and 0.35, respectively.
The results of these calculations are presented in Figure 9. For comparison, the figure also presents 
the results of the $UVO$ model, as well as a schematic presentation of the range of $\nu I(\nu)$ values 
allowed by the various observational
constraints and detections in the form of a shaded band. The detailed constraints were discussed earlier in Section
6.1 and presented in Figure 8. For the purpose of the discussion, we chose to represent the upper
limit in the 10 to 100
$\mu$m region by the value determined from the KMO96 fluctuation analysis, instead of the larger, but more definitive 
direct DIRBE limits (Paper I).

The results show that the $PFC$ model gives an acceptable fit to the EBL at $\lambda \gtrsim$ 100 $\mu$m. The
 discrepancies, the excess at $\lambda \gtrsim$ 400 $\mu$m and the deficiency at $\sim$ 140 $\mu$m,
 can be reduced by adopting a somewhat hotter spectrum for the emerging IR emission.
The $PFI$ model provides a better fit to the IR detections, but overestimates the local UV to near$-$IR
 luminosity density (see Table 2). However, the local luminosity density predicted by this model can be lowered if
 a significant fraction of the star formation produces massive stars, that recycle their 
energy instantaneously back to the ISM. 

To explore the predictions of such a two$-$component star formation scenario in more detail, we constructed a cosmic SFR
 that consists of the following sum:
\begin{equation}
\psi(z)=\psi_{_{UVO}}(z) + {\it C} \times\psi_{_{UVO}}(z)\ \left[ 1+{\psi_{_{UVO}}(z) \over \psi_0}\right]^{20} \quad ,
\end{equation}
where the first term represents the UVO cosmic SFR, and the second term represents an additional star formation 
rate component that produces massive stars. 
The latter component may be responsible for the observed iron enrichment
 in the intra$-$cluster medium (Elbaz, Arnaud, \& Vangioni$-$Flam (1995).
By construction, the functional form of the second component 
is sharply peaked  at the maximum of $\psi_{_{UVO}}(z)$, $\psi_0$ = 1 yr$^{-1}$ Mpc$^{-3}$, and
 {\it C} is a normalization constant, adjusted to provide an improved fit to the detected IR background.
 Figure 3 depicts the behavior of this SFR 
(hereafter refered to as $ED$) as a function of redshift, for {\it C} = $4.8\times 10^{-3}$. To examine the possibility that 
the cosmic SFR was significantly higher at redshifts above $\sim$ 1.5, as may be implied by the {\it ISO} observations
 of the HDF (RR97), we constructed an additional two$-$component cosmic SFR similar in form to the one given in eq. (27). 
In this model (hereafter designated $RR$),
 the first component is equal to $\psi_{_{UVO}}(z)$, and the second component is equal to 
$1.9\times 10^{-3}\ \psi_{_{UVO}}(z)\ \left[ 1+\psi_{_{UVO}}(z) / \psi_0\right]^{20}$ at redshifts $\lesssim 1.5$, and
 constant at its peak value at higher redshifts. The behavior of this SFR is also depicted in Figure 3. 
We assume that stars formed by the second component release all their radiative output instantaneously into the ISM, where it is 
totally absorbed by dust. The contribution of the second components to the luminosity density at various redshifts 
 is shown in Figure 4. 
The observed excess IR background emission, over that predicted by the various models shown in Figure 8, requires the rest frame spectra 
of these dusty, optically opaque, galaxies or star forming regions to peak shortwards of $\sim$ 100
$\mu$m. We therefore further assume that these opaque sources have a ULIRG spectrum, characterized by that of an H~II region (Fig.
6). The EBL spectrum predicted by this two$-$component SFR is therefore equal to the sum of that predicted by the $UVO$ model plus
 the cumulative redshifted IR emission arising from the second SFR component.

The resulting EBL spectra predicted by the two two$-$component models are shown in Figure 9.  Both models provide a significantly improved fit to
 the IR background compared to {\it all} other model predictions, and by construction, produce the local luminosity density
 at IR and UV$-$optical wavelengths. The $RR$ model
 predicts a somewhat larger IR background above $\sim$ 200 $\mu$m than the $ED$ model, since a larger fraction of its
 star formation rate takes place at higher redshifts.   
The $ED$ model requires the luminosity density at $z\approx$ 1.5 to be higher than that predicted by the $UVO$ model
 by about $\sim 2\times10^9$ L$_{\odot}$ Mpc$^{-3}$ (Fig. 4). If these sources are identified with 
ULIRGs having L
$\approx$ 10$^{13}$ L$_{\odot}$, then the required number density of such galaxies would be $\sim 10^{-4}$
Mpc$^{-3}$, about 10$^4$ times higher than that observed in the local universe. Current observations suggest only a mild
evolution in the ULIRG population up to $z \lesssim 0.1$ (e.g., Ashby et al.
1996; see also Sanders \& Mirabel 1996). Extrapolated to $z\ \approx\ 1$, this evolutionary trend yields an enhancement
in the ULIRG number density by a factor of only $\sim 100$ over the local value. Alternatively,
the excess emission could take place in dust$-$enshrouded starburst regions in a population of optically faint, but
 more numerous, galaxies 
that may have since dispersed their dust and evolved into the more familiar objects seen today.

Figure 9 also shows the contribution of several unobscured ``exotic" 
sources to the EBL (BCH86, BCH91). As discussed in Paper I, the contribution of some sources to the EBL can already be ruled
 out from the DIRBE upper limits in the 1.25 to 4.9 $\mu$m wavelength region. Complete obscuration of these sources by dust would 
move their spectra horizontally to longer wavelengths,
 with the location of the peak determined by the absorbing dust temperature. With some modification, these sources could contribute
 to the far$-$IR spectrum of the EBL. 
The possibility that some exotic and non$-$nuclear energy sources may contribute to the EBL, and the fact that the $ED$ and $RR$
 star formation histories produce similar IR background spectra shows that there is
 no unique way to account for the EBL spectrum. The current ambiguity in the magnitude and evolution of the cosmic star formation 
rate can only be resolved by future measurements of high$-z$ systems at infrared wavelengths.

\section{CONCLUSIONS} 
In this paper we examined some cosmological implications of the DIRBE and FIRAS detections of the EBL at
wavelengths between 140 to 5000 $\mu$m. 
We first showed that the residual isotropic infrared background is not likely to arise from local sources in the solar system or in the
Galaxy (\S 2). The DIRBE 140 and 240 $\mu$m and FIRAS 125$-$5000 $\mu$m residuals are therefore of extragalactic origin. Assuming that the 140 and
240 background radiation arises from dust emission, we used the measured intensities to derive a lower limit of 5 nW m$^{-2}$
sr$^{-1}$ for the extragalactic contribution at 100 $\mu$m.

The integrated EBL intensity detected by {\it COBE} in the 140 $-$ 1000 $\mu$m wavelength region is
$\sim$ 16 nW m$^{-2}$ sr$^{-1}$. This intensity is consistent with the energy release expected
from nuclear energy sources, and constitutes about 20$-$50\% of the total energy released in the formation of He and
metals throughout the history of the universe, the range reflecting various cosmic star formation histories considered in \S 4. 
Galaxy number counts provide a lower limit of 12 nW m$^{-2}$ sr$^{-1}$ on the EBL intensity
 in the 0.36 to 2.2 $\mu$m wavelength interval. The explored regions of the EBL account therefore for
 a total intensity in excess of 28 nW m$^{-2}$ sr$^{-1}$. Attributed only to nuclear sources, this intensity implies that more than 
$\sim$10\% of the baryonic mass
density implied by Big Bang nucleosynthesis analysis has been processed in stars to He and heavier elements. 

We examined the constraints set by the total intensity of the EBL detected by the {\it COBE} on the cosmological star formation rate (SFR), 
in particular, on the 
recently published rate compiled by MPD97 from UV and optical (UVO) observations. The EBL intensity predicted by this cosmic SFR is about 30
 nW m$^{-2}$ sr$^{-1}$, comparable to the EBL intensity in the 0.36$-$2.2 $\mu$m and 140$-$5000 $\mu$m wavelength regions. This
 leaves no room for any expected EBL in the $\sim$ 2 $-$ 140 $\mu$m wavelength interval. This suggests that the UVO cosmic SFR
missed a significant fraction of the 
cosmic star formation activity that takes place in dust$-$enshrouded galaxies or star forming regions. 

We constructed a detailed model for calculating the spectrum of the EBL using a hybrid of Forward, Backward, and Cosmic Chemical
 Evolution models. We calculated the EBL spectrum predicted for the UVO cosmic SFR, and compared this spectrum, as well as those 
predicted by other 
selected models, to the observational constraints (Figure 8). The 
IR background predicted by the UVO star formation rate falls considerably short of the {\it COBE} detections.
We therefore examined various alternative cosmic star formation histories. In particular we constructed a two$-$component 
cosmic SFR consisting of 
the UVO star formation rate plus an additional component representing the star formation activity that might be going
 on in dust enshrouded galaxies or star forming regions. Two such models, characterized by different behavior
 of the excess star formation activity at high redshifts were considered (see Figures 3 and 4). The EBL spectra predicted by these two models 
are very similar, and provide 
significantly better fits to the nominal {\it COBE} detections (Figure 9). The excess stellar energy must be generated mainly by 
massive stars which may be the source of the observed metal enrichment of the intra$-$cluster medium. 

Although there is currently no compelling need to invoke non$-$nuclear energy sources to explain the {\it COBE}
data, their potential contribution to the observed EBL cannot be ruled out. However, the magnitude and spectral
 shape of any such proposed contribution must now be consistent with the {\it COBE} upper limits and detections.
Their contribution, as well as the nucleosynthetic contribution to the EBL, cannot be unambiguously determined 
from the {\it COBE} data alone, and must await the results of future observations that will resolve the sources of the EBL.  
       
\acknowledgments
We acknowledge helpful discussions with Jim Felten, Jonathan Gardner, Alexander Kashlinsky, Simon Lilly,
Piero Madau, Bill Reach, Dave Sanders, and Vic Teplitz. We thank Alberto Franceschini, Bruno Guiderdoni, and Matt Malkan, for
communicating their model calculations, and Piero Madau for providing his updated cosmic SFR. We
thank Rebecca Bernstein for communicating her optical data prior to
publication. 
We thank Steve Shore and the referee Adolf Witt for their enthusiastic
response to the paper, and for suggestions that led us to consider more detailed modeling 
of the EBL spectrum. The authors also gratefully
acknowledge the contributions over many years of the many engineers, managers, scientists, analysts, and programmers engaged
in the {\it COBE} investigation. The National Aeronautics and Space Administration/Goddard Space Flight Center (NASA/GSFC)
was responsible for the design, development, and operation of the {\it COBE}. The GSFC was also responsible for the development of
the analysis software and for the production of the mission data sets. Scientific guidance is provided by the {\it COBE}
Science Working Group.

\appendix

\section{DISRUPTIVE FORCES AND INTERACTION TIMESCALES FOR A HYPOTHETICAL SOLAR SYSTEM CLOUD}

In the following we examine the lifetime of a hypothetical cloud of particles in the solar system, with properties constrained
 to produce an isotropic
 far$-$IR emission component comparable to the observed DIRBE residuals, 
against various disruptive processes in the solar system and the ISM. A similar study was conducted by Stern (1990), who considered the physical
interaction between the ISM and the Oort cloud. To set strict upper limits on the stability of this hypothetical cloud we will
assume that it consists of refractory particles with a mass density $\rho$ = 3 g cm$^{-3}$. 

\subsection{Particle Cloud Interaction with the Interstellar Medium}
Interactions with charged constituents of the ISM are clearly important if the particle cloud is located outside the
heliosphere, the cavity carved out from the ISM by the solar wind. The radius of this cavity is determined by the distance
at which the magnetic pressure of the solar wind equals that of the ISM, and could be anywhere between 50 and 150 AU (Holtzer
1989). We will assume that interactions with the ISM can also take place within the heliosphere, since the ULYSSES spacecraft
has detected a flux of interstellar dust particles in the solar system (Gr\"un et al. 1994), and neutral atoms can penetrate
the heliosphere in a similar fashion.

The solar system moves at a velocity
$v_{\odot} \approx$ 20 km s$^{-1}$ relative to the ISM gas (Spitzer 1978). This motion will take the solar system through 
various ISM phases characterized by different densities ($n$), temperatures ($T$), and filling factors ($f$): the hot
ionized medium (HIM) characterized by
\{$n_h, T_h, f_h$\} =
\{0.005 cm$^{-3}$, 10$^6$ K, 0.2$-$0.6\}; the warm neutral medium (WNM) characterized by \{$n_w, T_w,
f_w$\} =
\{0.26 cm$^{-3}$, 8000 K, $\sim$ 0.3\}; and the cold neutral medium (CNM) characterized by \{$n_c, T_c, f_c$\} =
\{40 cm$^{-3}$, 50 K,
$\sim$ 0.02\}. The average ISM density traversed by the solar system is therefore $n_0 \sim$ 1 cm$^{-3}$.

Thermal sputtering will be most important when the solar system passes through the HIM. The sputtering
lifetime of a refractory dust particle of radius $a$ embedded in a gas of number density $n$ and temperature T
$\approx 10^6$ K is (Dwek, Foster, \& Vancura 1996):

\begin{equation}
\tau_{sput} \approx 5\times 10^6 {a(\mu \hbox{m})\over n(\hbox{cm}^{-3})}\ \ \ \ \hbox{yr}
\end{equation}

\noindent
Adopting a HIM filling factor of 0.4, the solar system has spent a total of $\sim$ 1.8$\times 10^9$ yr in
the HIM. Any dust particles with radii $\lesssim$ 1.8 $\mu$m would therefore have been eroded away by thermal
sputtering during this time period.

Drag will become important when a particle has swept up the equivalent of its own mass. The drag time is
therefore given by:

\begin{eqnarray}
\tau_{drag} &  = & {4 \rho a \over 3 n_0 v_{\odot} \mu m_H\ } \nonumber \nl
            & = & 2.7\times 10^{10}\ a\hbox {(cm) yr}
\end{eqnarray}

\noindent
where we used $\rho
\approx 3\ $g cm$^{-3}$ for the average density of a particle, and
$\mu \approx$ 1.4 as the mean molecular weight (in amu) of the ISM gas. Dust particles with radii
$\lesssim 0.16$ cm will therefore be lost from the cloud by collisional drag with the ISM on a timescale
$\lesssim \tau_{\odot}$.

Kinetic sputtering is important when dust particles move at suprathermal speed through the ISM. The timescale
for destruction by kinetic sputtering following a collision with ISM gas atoms with number density $n_A$ and
 mass $m_A$ is

\begin{equation}
\tau_{kin} = {4 \rho a \over 3 n_A v_{\odot} \mu_{sp} m_H\ Y(E)},
\end{equation}

\noindent
where $\mu_{sp}\approx 20$ is the atomic mass of a sputtered particle, and $Y(E)$ is the sputtering yield
(defined as the number of ejected particles per incident projectile) at energy $E\ =\ {1\over 2} m_A\ v_{\odot}^2$. At
velocities of 20 km s$^{-1}$, H, He, and C atoms will impinge on the cloud particles with kinetic energies of 2, 8, and 24
eV, respectively. 
Comparison of the drag and kinetic sputtering timescales gives:

\begin{equation}
{\tau_{drag}\over \tau_{kin}} = {n_A \mu_{sp} Y(E) \over n_H \mu}. 
\end{equation}

\noindent
Sputtering thresholds range from
$\sim$ 5 to 8 eV for the most common refractory material, so that only sputtering by He or C atoms will be
important, giving $n_A \mu_{sp}$/$n_H \mu \sim$ 1. Furthermore, close to the sputtering threshold, $Y(E) \ll 1$.
Consequently, cloud disruption by kinetic sputtering can be neglected compared to its erosion by collisional drag.

Evaporative collisions with interstellar grains are not an important destruction process for the cloud
particles. Laboratory experiments on collisions between refractory rocks (O'Keefe \& Ahrens 1982) show that
complete vaporization occurred only when the kinetic energy per projectile atom exceeded the material binding
energy (typically 5 - 8 eV) by factors $\gtrsim$ 5. These correspond to impact velocities $\gtrsim$ 30 km
s$^{-1}$, well above the relative velocity of the solar system particle cloud with the ISM. 
However, the slow and continuous
erosion of the cloud particles by cratering collisions proceeds at much lower impact velocities, and can be
an important destruction process. Following Borkowski \& Dwek (1995) we write the erosion rate due to
cratering collisions as:

\begin{equation}
\left({dm\over dt}\right)_{cr} =  \pi a^2 n_d(m_d)\ v_{\odot}\ m_A\ Y_{cr}(E)
\end{equation}

\noindent
where $n_d(m_d)$ is the number density of interstellar dust grains (i.e. projectiles) of mass $m_d$, $m_A$ is the
 mass of an individual atom in the target material, and $Y_{cr}=m_d\
v^2_{\odot}/2E_{cr}$ is a dimensionless cratering yield, where $E_{cr}$ is the specific energy for the ejection of
one atom in a cratering event, typically $\sim$ 1 eV. The product $m_A Y_{cr}$ is thus equal to the total mass
excavated in the cratering event, and the validity of the expression requires it to be significantly less than
$m$, the mass of the target grain. The timescale for particle erosion by cratering collisions is simply given by
$m/\dot m_{cr}$:

\begin{eqnarray}
\tau_{cr} &  = & {4 \rho a \over 3  n_0 v_{\odot}\ Z_d \mu m_H} 
                     \left({m_A v_{\odot}^2\over 2E_{cr}}\right)^{-1} \nonumber \nl
    &  = & 6.8\times 10^{10}\ a(cm)\ \ \ \  yr
\end{eqnarray}

\noindent
where we used the relation: $n_d(m_d) m_d = Z_d n_0 \mu m_H $, with $Z_d = 0.01$ as the dust to gas mass
ratio in the ISM,  $m_A = 20\ m_H$, and $E_{cr}$ = 1 eV. Equation (9) is valid only for particles with
 $m \gtrsim 40\ m_d$, or $a \gtrsim$ 4 \mic, where we adopted a grain radius of $\sim$ 1 $\mu$m for the largest
interstellar dust particles. Hence cloud particles with sizes less than $\sim$ 0.07 cm will be eroded by
cratering collisions with interstellar dust particles.

\subsection{Forces Within the Solar System}
Particles within the heliosphere will also be subjected to local forces caused by the Sun's gravity, its
radiation and the solar wind.
Depending on the optical properties, mass, and heliocentric distance of the particles, they can spiral into the
Sun, be blown out of the solar system, or remain unaffected by the combined effect of these forces.
The absorption and reradiation of sunlight by the orbiting particles gives rise to a 
Poynting$-$Robertson (P$-$R) drag.
The decay time for heliocentric orbits of radius $d$ as a result of the P$-$R drag is given by Burns, Lamy, \&
Soter (1979; hereafter BLS79):
\begin{eqnarray}
\tau_{P-R} & = & {mc^2\over 4 F_{\odot}(d)\ \pi a^2 Q_{pr}} \nonumber \nl
            &  = & 7\times 10^2\ {\rho\ a(\mu \hbox{m})\ d^2(\hbox{AU})\over Q_{pr}}\ \ \ \ \hbox{yr}
\end{eqnarray}

\noindent
where $F_{\odot}(d)$ is the solar bolometric flux at distance $d$, and $Q_{pr}$ is the solar$-$spectrum$-$averaged radiation
pressure efficiency factor of the particle (see below). A lower limit on the radius of the particles that will survive the
P$-$R effect can be obtained by adopting a large heliopause radius of 150 AU. At this distance the critical radius above
which particles will maintain a stable orbit over the lifetime of the solar system is given by $a_{crit}(\mu \hbox{m}) = 95\
Q_{pr}$.

Radiation pressure can also expel particles from the solar system, if it overcomes the Sun's gravitational attraction.
Following BLS79, we define the parameter $\beta$ as the ratio between the two forces:

\begin{equation}
\beta \equiv {F_{rad}\over F_{grav}} = 0.57 {Q_{pr}\over \rho\ a(\mu m)}
\end{equation}

\noindent
Particles with $\beta \geq$ 1/2 can escape the solar system (BLS79). Grains with radii of $a_{exp}(\mu \hbox{m}) \leq 0.09\
Q_{pr}$ can therefore be expelled from the solar system by radiation pressure. These radii are
smaller than
$a_{crit}$, so particles that are stable against the P$-$R effect are also stable against expulsion from the solar system by
radiation pressure.

The radiation pressure efficiency factor of the particle is given by
$Q_{pr}=Q_{ext}[1-<\cos(\theta)>\ A]$, where $Q_{ext}$ is the extinction efficiency of the particle, $A$
its albedo, and $<\cos(\theta)>$ is a measure of its forward scattering efficiency.
For a strongly forward scattering particle $<\cos(\theta)>\ \approx 1$, whereas for an isotropically
scattering particle $<\cos(\theta)>\ \approx 0$. Particles with radii significantly greater than the
wavelength of the incident radiation are strong forward scatterers, and applying geometrical optics one
gets a value of $Q_{pr} \approx $0.5, and $Q_{ext} \approx$ 1.

\subsection{Gravitational Perturbations from Passing Stars}

Any cloud of particles cannot survive beyond the tidal radius $R_T$, at which the Galactic tidal forces on a
particle are equal to the Sun's gravitational force. The tidal radius is given by (Tremaine 1993):

\begin{equation}
R_T(\hbox{AU}) = 1.7\times10^5 \left({\rho_{gal}\over 0.15\ \hbox{M}_{\odot}\ \hbox{pc}^{-3}}\right)^{-1/3}
\end{equation}

\noindent
where $\rho_{gal}$ is the local Galactic mass density.
Outside this radius, cloud particles can be
ejected from the solar system by gravitational perturbations from passing stars and other material. The half life for this
process is given by (Weinberg, Shapiro, \& Wasserman 1987; Tremaine 1993):

\begin{equation}
\tau_{gal} =  10^{14}\  d^{-1}(\hbox{AU}) \left({\rho_{gal}\over 0.15\ M_{\odot}\ \hbox{pc}^{-3}}\right)^{-1}\ \ \ \ yr
\end{equation}

\noindent
Cloud particles located at distances above $2\times 10^4$ AU will therefore be stripped out of the solar system.



\begin{deluxetable}{cccccc}
\small
\tablecaption{Intensity of Interplanetary Components}
\tablehead{
\colhead{Component}&
\colhead{Distance}&
\colhead{Temperature}&
\colhead{Mass}&
\colhead{$\nu I_{\nu}$(240\micron)}&
\colhead{Isotropy}\\
\colhead{}&
\colhead{(AU)}&
\colhead{(K)}&
\colhead{(g)}&
\colhead{(nW m$^{-2}$ sr$^{-1}$)}&
\colhead{}\\
}
\startdata
Oort Cloud        & $2\times10^4-10^5$ & 5        & $4\times10^{28}$        &
$8\times10^{-6}$ & yes \nl
Inner Oort Cloud  & $10^3-2\times10^4$ & 10       & $2\times10^{29}$        &
$1\times10^{-3}$ & yes \nl
Kuiper Belt       & $40-100$           & 40       & $<6\times10^{27}$       &
$0.02$ & no \nl
Kuiper Belt Dust  & $40-100$           & 40       & $6\times10^{27}$        &
$6\times10^{4}$ & no \nl
Asteroidal Bodies & $1.8-3.8$          & 180      & $2\times10^{22}$        &
$0.04$ & no \nl
Meteoroids        & $<10$              & 100--275 &        $10^{19}$        &
$0.005$ & no \nl
Interstellar Dust & $\ge2$             & $<$200   & $2\times10^{-21}/$m$^3$ &
$0.004-0.02$ & yes \nl
Zodiacal Light    & $\le3.5$           & 275      & $10^{19}-10^{20}$       &
$25$ & no \nl
Hypothetical Cloud           & 60                 & 30   & $10^{26}(\frac{a}{\rm cm})$ &
$14$\tablenotemark{a} & yes \nl
\enddata
\tablenotetext{a}{Assumed equal to the residual isotropic emission (Eq. 1 and Fig 1a).}
\end{deluxetable}


\begin{deluxetable}{lllll}
\tablecaption{Select Model Output Parameters\tablenotemark{a}}
\tablehead{
\colhead{  Quantity } &
\colhead{ UVO } &
\colhead{  PFI } &
\colhead{ PFC} &
\colhead{ Observed} \\
 }
\startdata 
$I$ [nW m$^{-2}$ sr$^{-1}$]     & 30 & 91 & 41 &  {\it 28} \tablenotemark{b}  \nl
$\epsilon(z=0)$ [L$_{\odot}$ Mpc$^{-3}$] & $1.9\times10^8$ & $4.1\times10^8$ & $1.9\times10^8$ &  
       ${\it (1.8\pm 0.7)\times10^8}$ \tablenotemark{c}  \nl
$\epsilon(4400\AA, z=0)$ [W Hz$^{-1}$ Mpc$^{-3}$] & $2.0\times10^{19}$ & $3.2\times10^{19}$ & $1.8\times10^{19}$ &  
       ${\it (2.0\pm 0.8)\times10^{19}}$ \tablenotemark{d}  \nl
$\epsilon_{IR}(z=0)$ [L$_{\odot}$ Mpc$^{-3}$] & $5.0\times10^7$ & $6.3\times10^7$ & $4.5\times10^7$ & 
     ${\it 5.3\times10^7}$ \tablenotemark{c}  \nl
$\tau$(V)                             & 0.27 & 0.20 & 0.35 &  \nodata \nl

\enddata
\tablenotetext{a}{models are explained in \S 4 of the text}
\tablenotetext{b}{observed in the 0.36$-$2.2 and 140$-$5000 $\mu$m wavelengths interval}
\tablenotetext{c}{see \S 5.2 of the text}
\tablenotetext{d}{Lilly et al. (1996)}  
\end{deluxetable}


\begin{deluxetable}{lllll}
\tablecaption{Dust Emission Components of {\it IRAS} Galaxies and their Average Flux Ratios\tablenotemark{a}}
\tablehead{
\colhead{  log$\left({L_{IR}\over L_{\odot}}\right)$ } &
\colhead{  $f$(H II) } &
\colhead{  $f$(H I)  } &
\colhead{  ${S_{\nu}(12\ \mu\hbox{m})\over S_{\nu}(25\ \mu \hbox{m})}$ } &
\colhead{  ${S_{\nu}(60\ \mu\hbox{m})\over S_{\nu}(100\ \mu \hbox{m})}$ } \\
 }
\startdata 
$<$ 9.5 & 0.27 & 0.73 & 0.69 & 0.36  \nl
 9.75 & 0.28 & 0.72 & 0.68 & 0.37  \nl
10.25 & 0.34 & 0.66 & 0.59 & 0.42  \nl
10.75 & 0.45 & 0.55 & 0.46 & 0.51  \nl
11.25 & 0.61 & 0.39 & 0.32 & 0.66  \nl
11.75 & 0.78 & 0.22 & 0.22 & 0.84  \nl
12.0 & 0.88 & 0.12 & 0.17 & 0.96  \nl
 13.0 & 1.0 & 0.0 & 0.13 & 1.11 \nl
\enddata
\tablenotetext{a}{$f$(H II) and $f$(H I) are, respectively, the luminosity$-$dependent fractional contribution of the H~II
and H~I dust spectra (shown in Figure 5) to the spectra of the {\it IRAS} galaxies. The flux ratios in columns 4 and 5 are the
derived flux ratios of the galaxies, and should be compared with the mean observed values given in Table 5 of Soifer \&
Neugebauer (1991). The agreement is {\it exact} for the 12 to 25 $\mu$m flux ratio, and deviates from the
60 to 100 $\mu$m flux ratio by less than $\sim$ 10\%. Galaxies with L$_{IR} = 10^{13}$ L$_{\odot}$ are represented by a
pure H~II spectrum.}
\end{deluxetable}

\clearpage

\begin{figure}
\caption{Spectra of hypothetical clouds of particles that can contribute
significantly to the residual emission at 140 and 240 $\mu$m within the {\it COBE} observational constraints. 
Filled circles represent the DIRBE upper limits and detections (Paper I), and the open diamonds represent the 
FIRAS dark sky limits (Shafer et al. 1997).
Cloud particles radiate with a
$\nu^n$ emissivity law. The upper panel shows the constraints for $n = 0$ particles. Particle temperatures are constrained to
be below 100 K by the DIRBE upper limits at wavelengths $\leq$ 100 $\mu$m, and above 18 K by the FIRAS dark sky limits.
Particles with a temperature of 30 K can, in principle, produce {\it all} the 140 and 240 $\mu$m residual emission in these
bands. Also shown in square brackets, are the heliocentric distances and masses calculated for such clouds, if they were 
located in the solar system. The lower panel shows the constraints for $n = 2$ particles. The lower
temperature limit is determined by the interstellar radiation field. }
\end{figure}

\begin{figure}
\caption{The size$-$distance \{$a$(cm), $d$(AU)\} parameter space that must be occupied by any interplanetary cloud of particles
that produces a significant isotropic signal in the DIRBE 140 and 240 $\mu$m bands. Shaded regions indicate regions in \{$a$,
$d$\} parameter space which are not stable over the lifetime of the
solar system or are excluded by the temperature constraints imposed by the
DIRBE data. The time scales for the various destructive
processes are summarized in Section 2 and the Appendix.  }
\end{figure}

\begin{figure}
\caption{The cosmic star formation rate, $\rho_*$(z), as a function of redshift. Shown are rates inferred from UV and optical
 observations, recently compiled by MPD97 (bold solid line; labeled UVO), and the rates inferred for 
various star forming scenarios in the cosmic chemical evolution model of PF95: the infall model (thin solid line; PFI), 
and the closed box model (dotted line; PFC). The long$-$dashed line (labeled ED) represents a 
cosmic star formation rate constructed to fit the EBL intensities detected by {\it COBE}. The short$-$dashed line (RR)
 represents a cosmic star formation rate with an excess of sources at high redshifts, similar to that
 suggested by RR97 to account for the excess IR galaxies observed by {\it ISO} in the HDF.}
\end{figure} 

\begin{figure}
\caption{The cosmic luminosity density, $\epsilon(z)$, calculated for the various star formation histories
 depicted in Figure 3. The $ED$ and $RR$ curves represent the {\it excess} luminosity density associated with the 
ED and RR star formation rates, over that of the UVO model (bold solid line).}
\end{figure}

\begin{figure}
\caption{The local spectral luminosity density at 0.28, 0.44, and 1.0 $\mu$m (Lilly et al. 1996), and at 2.2 $\mu$m
(Gardner et al. 1997) indicated by solid circles is compared to the starlight emitted by an average spiral galaxy (Schmitt et
al. 1997; dashed line). The heavy solid curve represents the average 
IR spectrum of {\it IRAS} galaxies. Its
derivation is described in \S 5.2 of the text.}
\end{figure}

\begin{figure}
\caption{The arbitrarily normalized spectral luminosity of the H II and H I dust components that were used to construct the
{\it IRAS} galaxy spectra. The H II spectrum represents a somewhat hotter version of the Orion spectrum obtained from the DIRBE
data (Wall et al. 1996). The diamonds depict the ISOPHOT and LWS spectrum of Arp 220 obtained with the {\it ISO} satellite (Klaas et al. 1996),
arbitrarily normalized for comparison with the H II spectrum. The figure shows that the H II spectrum is a reasonable
representation of that of interacting galaxies. The H I spectrum represents the model fit of Dwek et al. (1997) to the DIRBE
observations of the IR emission from the diffuse H I component of the Galaxy.}
\end{figure}

\begin{figure}
\caption{The spectral luminosity of representative {\it IRAS} galaxies listed in Table 3. The lowest curve represents that of
normal galaxies (L$_{IR}<3\times10^9$ L$_{\odot}$), whereas the highest curve represents that of the most luminous {\it IRAS}
galaxies (L$_{IR}>10^{13}$ L$_{\odot}$). }
\end{figure}

\begin{figure}
\caption{Model calculations of the EBL spectrum are compared with the {\it COBE} results. The DIRBE 2$\sigma$ upper limits at
1.25 $-$ 60 $\mu$m, and detections at 140 and 240 $\mu$m with
$\pm$ 2$\sigma$ error bars are represented by solid circles (Paper I). The 100 $\mu$m intensity is represented
by the 95\% confidence limit of 5$-$34 nW m$^{-2}$ sr$^{-1}$. The FIRAS 125$-$5000 $\mu$m detection
(Fixsen et al. 1997) is shown by a light dashed line; the UV$-$optical lower limits of Pozzetti et al. (1997) and the 2.2 $\mu$m
lower limit (Gardner et al. 1997) are represented by solid squares.  The ``X" represent the upper limits on the EBL derived from TeV
observations of Mrk 501 (Stanev
\& Franceschini 1997), and Mrk 421 (Dwek \& Slavin 1994), and the open diamonds represent the upper limits derived from the
analysis of the fluctuations the DIRBE maps (KMO96). The heavy solid curve represents the EBL spectrum calculated from the $UVO$ model 
in this paper. Other models presented in the figure are
the Backwards Evolution models of Malkan \& Stecker (1997; dotted line), and of Beichman \& Helou (1991; dashed dotted
line); the Forward Evolution models of Franceschini et al. (1997, dashed line), and Guiderdoni et al. (1997, dashed
triple dotted line); and the Cosmic Chemical Evolution model of Fall, Charlot, \& Pei (1996, thin solid line).
The dashed line marked $S$ represents the EBL spectrum calculated for a dust$-$free universe with the UVO cosmic star
formation rate.
 }
\end{figure}

\begin{figure}
\caption{The observational constraints from Figure 8 (represented here by a shaded area) are compared with the EBL calculated
for several different cosmic star formation histories. The DIRBE and
FIRAS detections are represented by the
$\pm$2$\sigma$ uncertainties in their values, and the DIRBE 2$\sigma$ upper limits are represented by solid circles. The 10 $-$ 100
$\mu$m upper boundary of the shaded area represents the KMO96 upper limit. The thin solid line represents the EBL predicted
 by our model using the UVO cosmic SFR (see also Fig. 8). 
The long$-$ and short$-$dashed curves represent the EBL predicted by our model using the PFI and PFC star formation rates, respectively. 
The two thick lines
 represent the EBL predicted if the UV$-$optically determined cosmic SFR is augmented by a ``hidden" component representing 
the star formation activity taking place in
dust$-$enshrouded galaxies or star forming regions.
The thick solid line represents the EBL if the excess luminosity density of these objects is distributed with redshift
 as the curve marked $ED$ in Figure 4, and the thick dashed line assumes an excess luminosity density that varies with redshift
 as curve marked $RR$ in Figure 4.
Also shown in the figure are the EBL spectra of various sources postulated by BCH86 and BCH91: exploding stars ($ES$), decaying
particles ($DP$), very massive objects ($VMO$), halo black holes ($HBH$), primeval galaxies ($PG$), and AGNs, all calculated for
H$_0$ = 50 km s$^{-1}$ Mpc$^{-1}$. The dotted line marked $S$ represents the EBL spectrum calculated for a dust$-$free universe with the UVO cosmic star
formation rate. }
\end{figure}
    
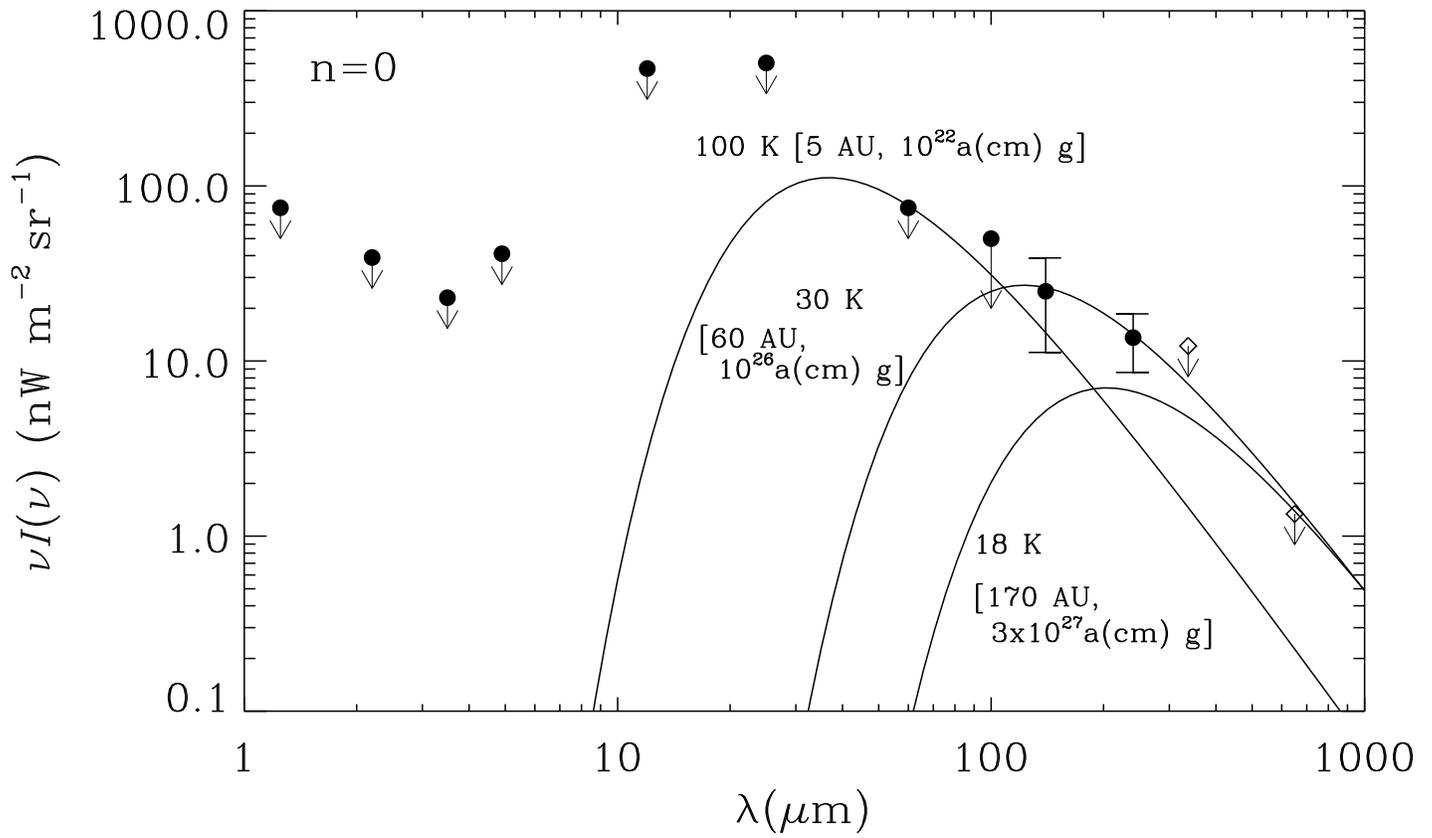
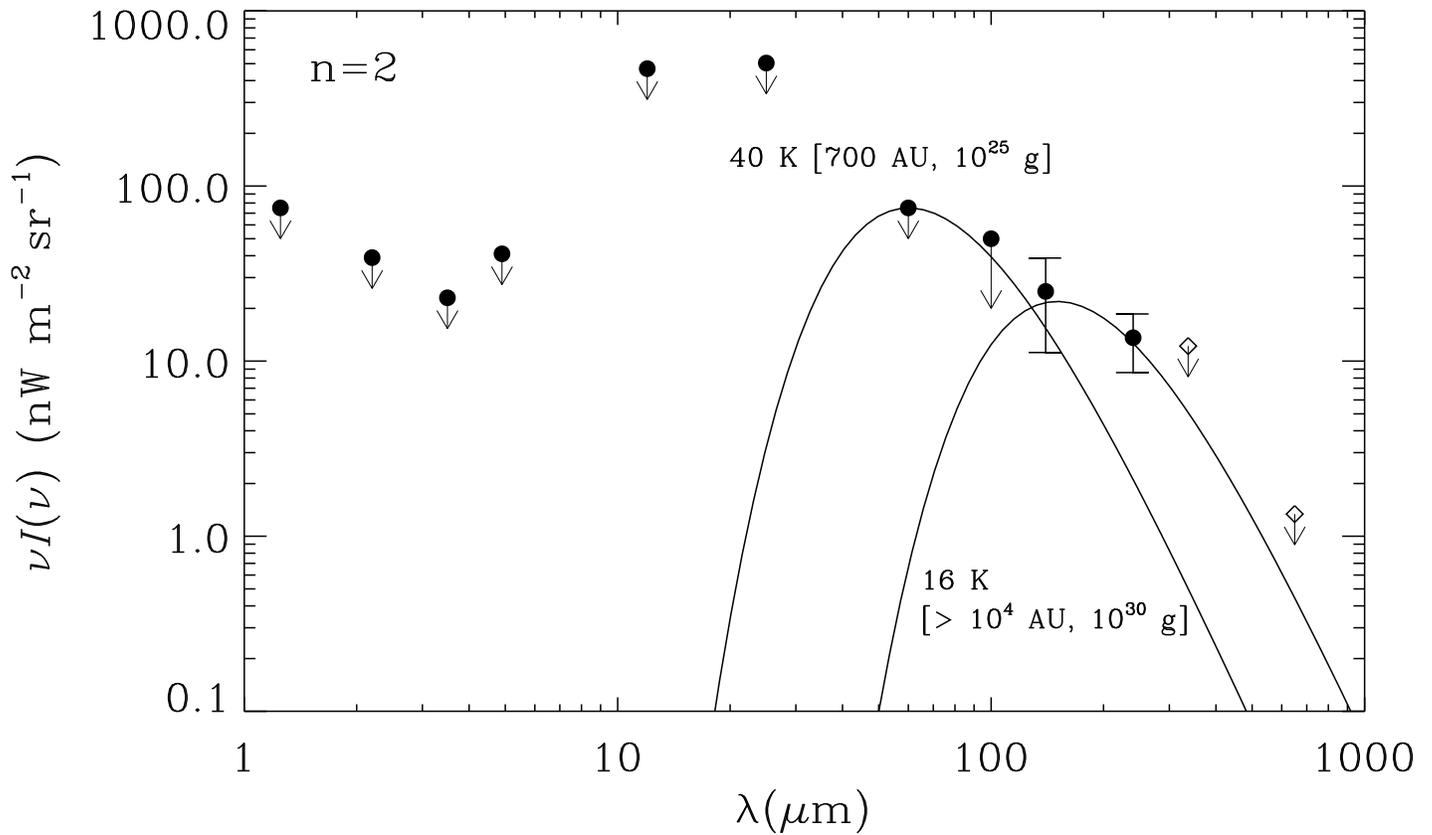
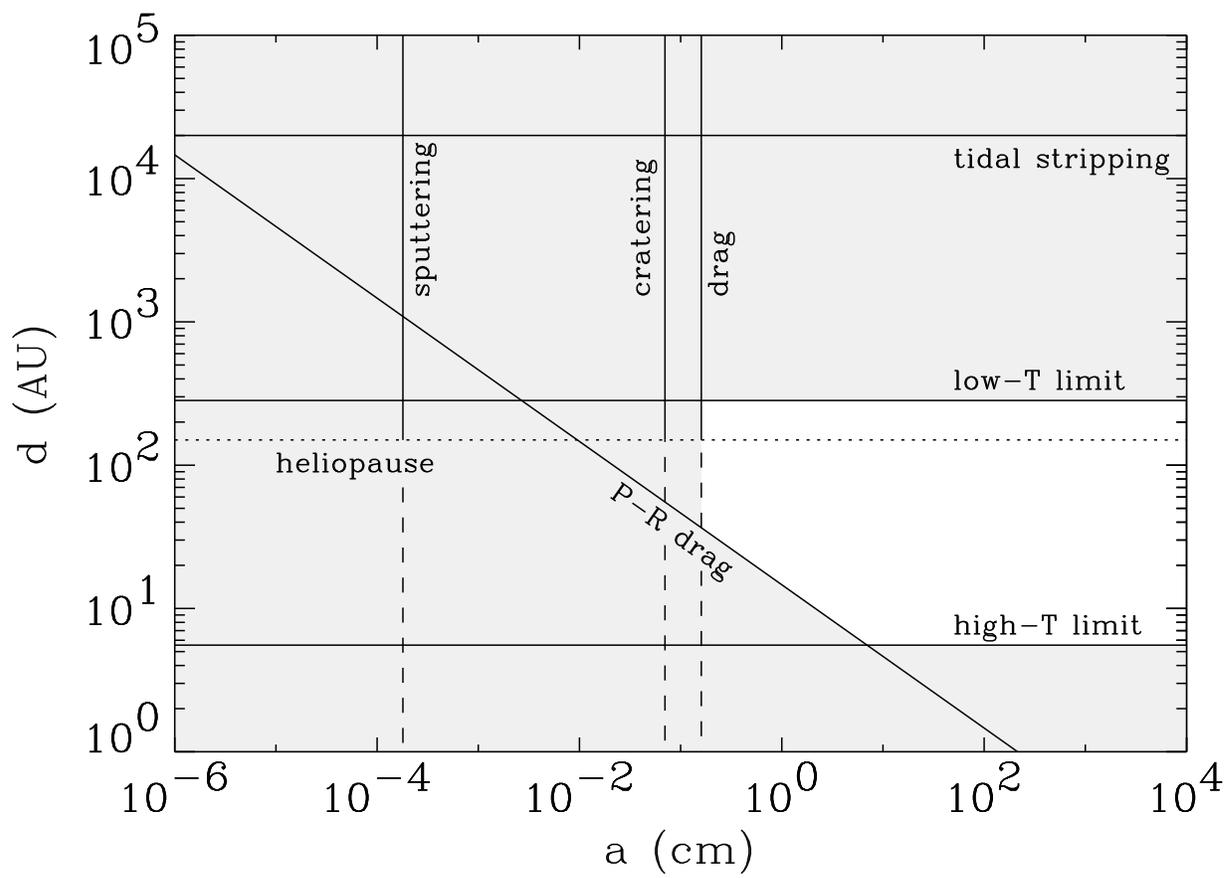
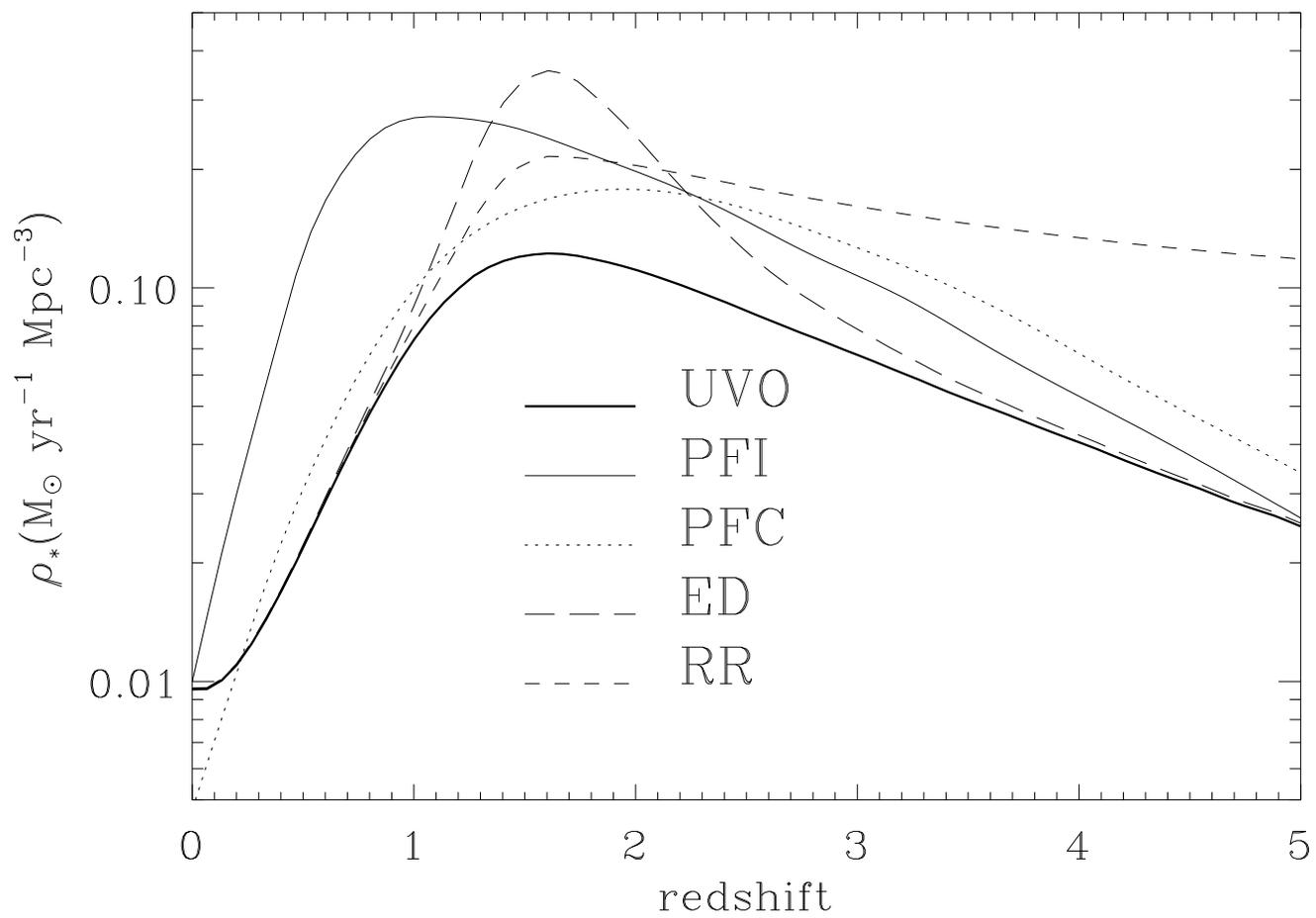
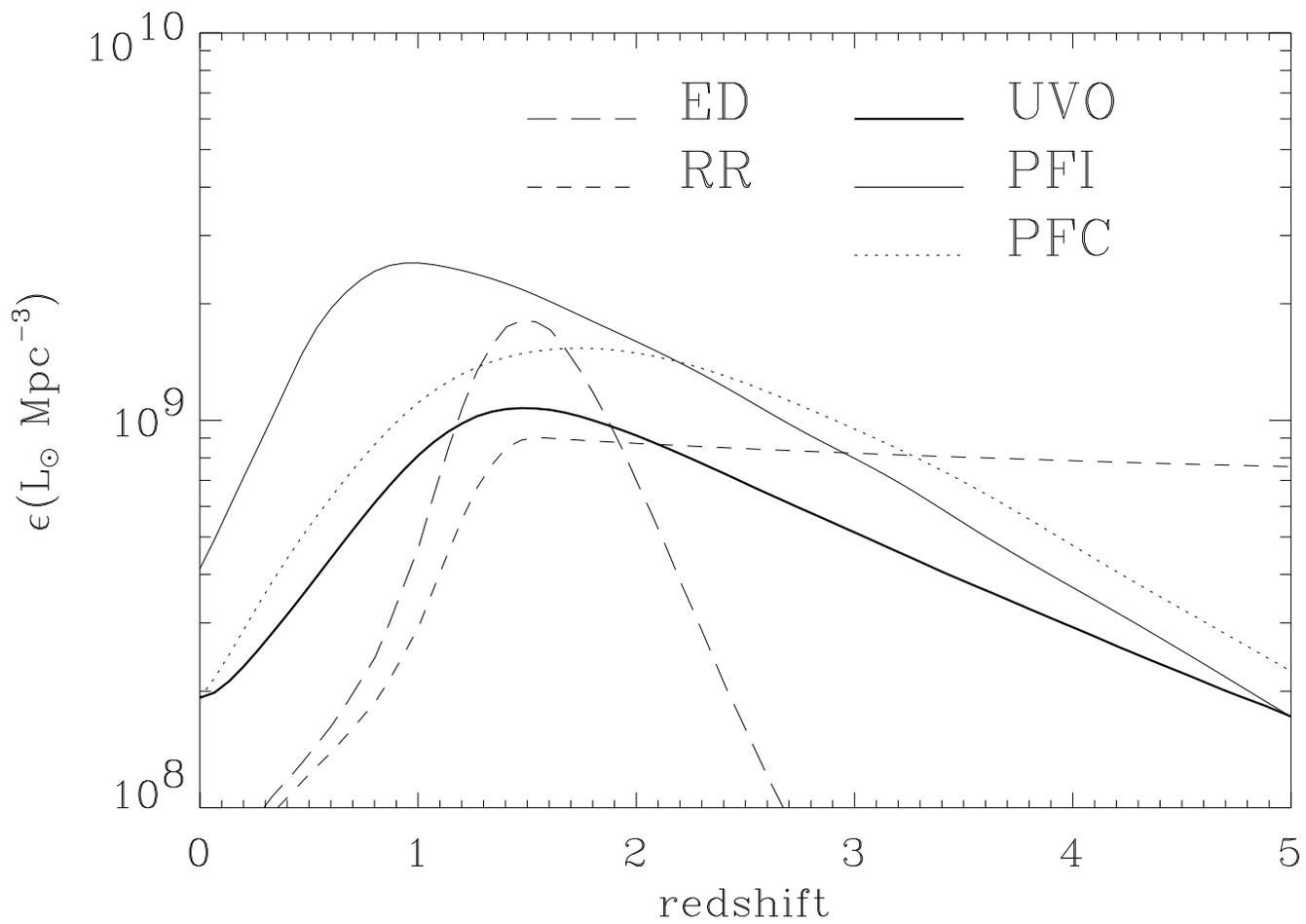
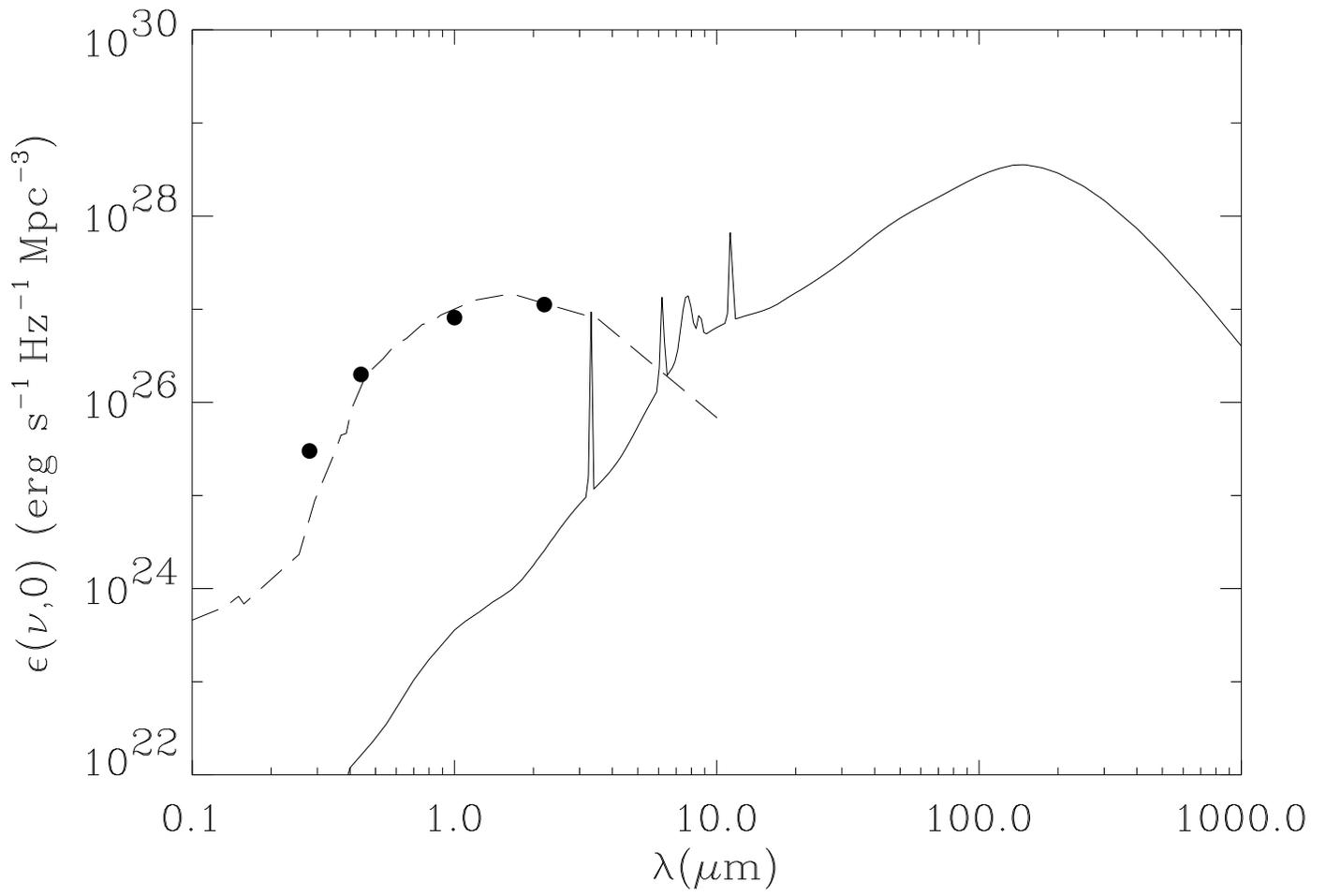
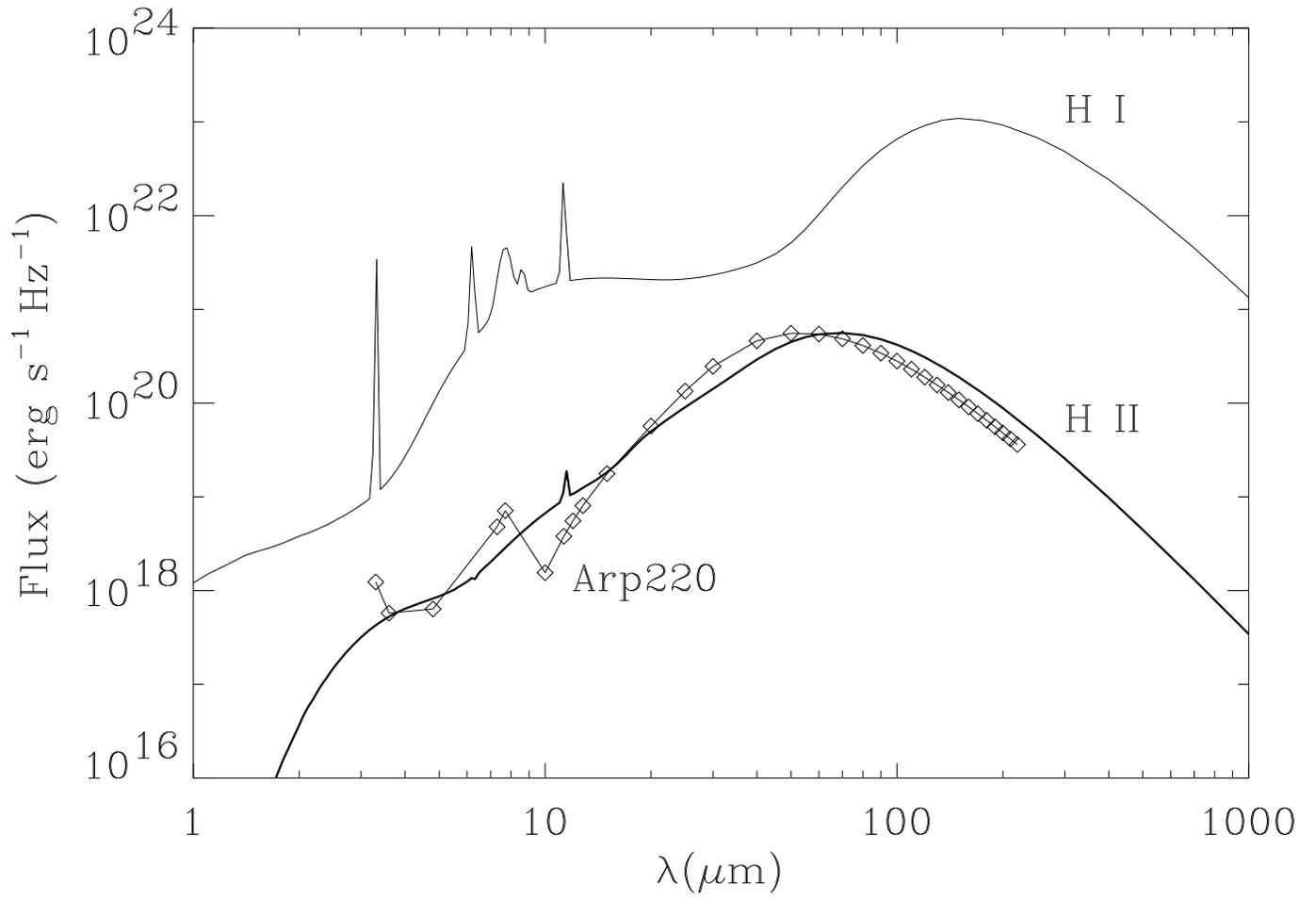
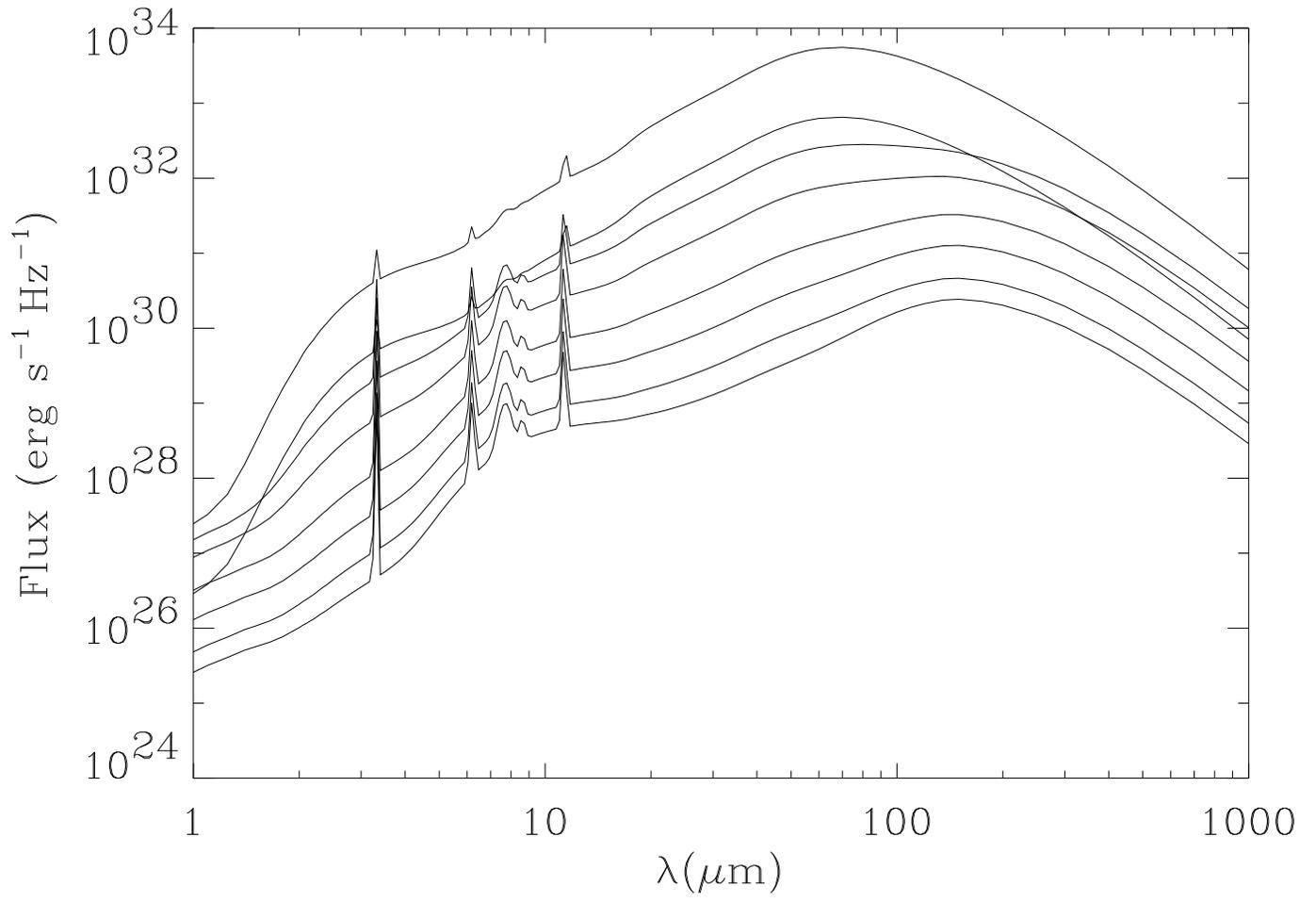
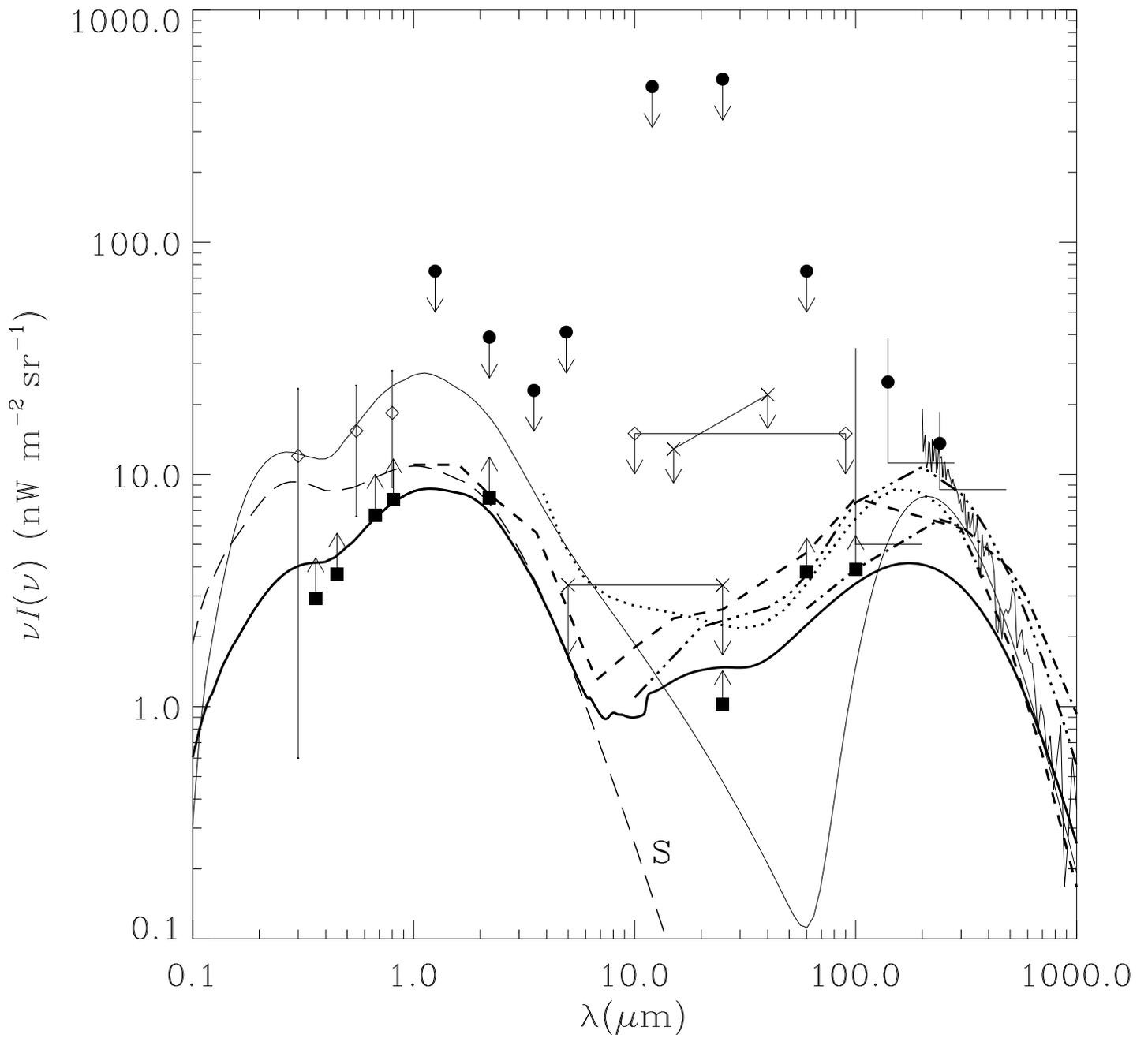
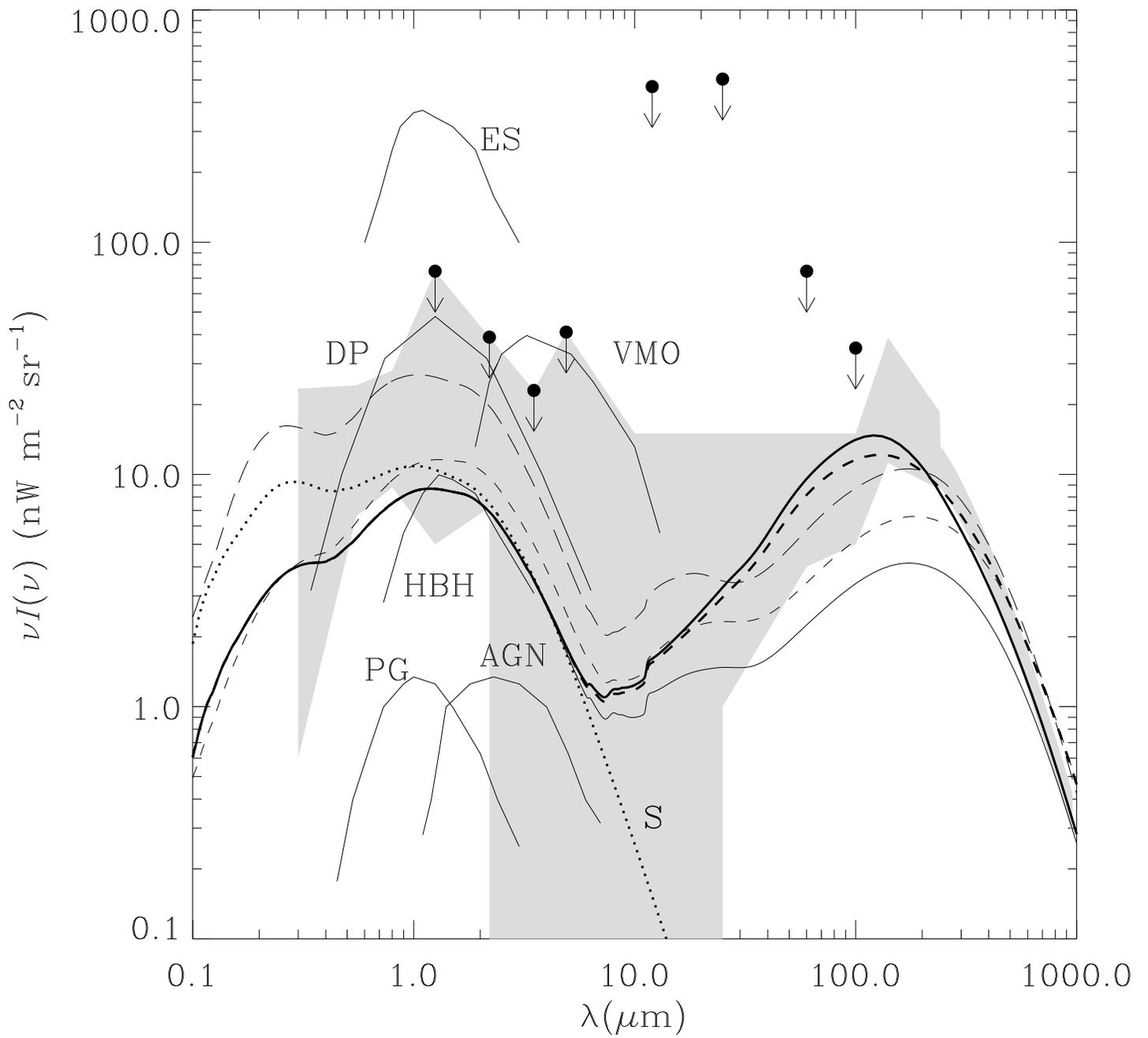
\end{document}